\documentclass[journal=jctcce,manuscript=article]{achemso}

\usepackage[version=3]{mhchem} 
\usepackage{hyperref}
\usepackage{braket}
\usepackage{bm}

\usepackage[authormarkuptext=name,authormarkup=none]{changes}
\definecolor{acolour}{RGB}{0, 0, 255}
\definecolor{bcolour}{RGB}{255, 0, 0}
\definechangesauthor[name={Yorick}, color={acolour}]{YS}
\definechangesauthor[name={Gianluca}, color={acolour}]{GL}

\usepackage{mathrsfs}
\DeclareMathAlphabet{\mathscrbf}{OMS}{mdugm}{b}{n}

\newcommand*{\matr}[1]{\mathbf{#1}}
\newcommand*{\vect}[1]{\bm{#1}}

\author{Yorick L. A. Schmerwitz}
\affiliation{
Science Institute and Faculty of Physical Sciences, University of Iceland, Reykjav\'{\i}k, Iceland
}
\author{Gianluca Levi}
\email {giale@hi.is}
\affiliation{
Science Institute and Faculty of Physical Sciences, University of Iceland, Reykjav\'{\i}k, Iceland
}
\author{Hannes J\'onsson}
\email {hj@hi.is}
\affiliation{
Science Institute and Faculty of Physical Sciences, University of Iceland, Reykjav\'{\i}k, Iceland
}

\title{Calculations of Excited Electronic States by Converging on Saddle Points Using Generalized Mode Following}

\begin{document}

\renewcommand*\tocentryname{TOC Graphic}
\begin{tocentry}
   \includegraphics[width = 0.99\textwidth]{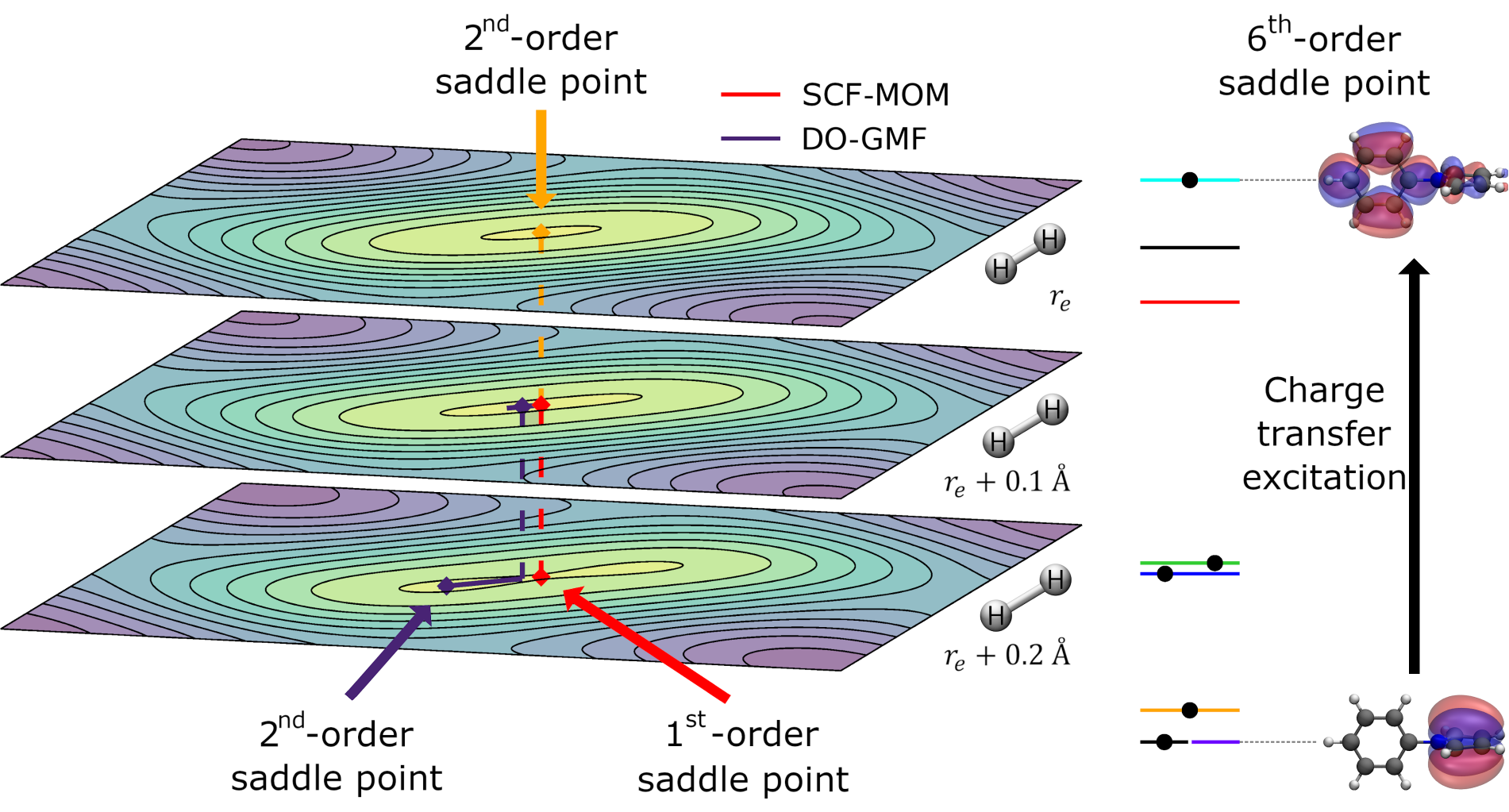}
\end{tocentry}

\begin{abstract}
Variational calculations of excited electronic states are carried out by finding saddle points on the surface that describes how the energy of the system varies as a function of the electronic degrees of freedom. This approach has several advantages over commonly used methods especially in the context of density functional calculations, as collapse to the ground state is avoided and yet, the orbitals are variationally optimized for the excited state. This optimization makes it possible to describe excitations with large charge transfer where calculations based on ground state orbitals are problematic, as in linear response time-dependent density functional theory. A generalized mode following method is presented where an $n$\textsuperscript{th}-order saddle point is found by inverting the components of the gradient in the direction of the eigenvectors of the $n$ lowest eigenvalues of the electronic Hessian matrix. This approach has the distinct advantage of following a chosen excited state through atomic configurations where the symmetry of the single determinant wave function is broken, as demonstrated in calculations of potential energy curves for nuclear motion in the ethylene and dihydrogen molecules. The method is implemented using a generalized Davidson algorithm and an exponential transformation for updating the orbitals within a generalized gradient approximation of the energy functional. Convergence is found to be more robust than for a direct optimization approach previously shown to outperform standard self-consistent field approaches, as illustrated here for charge transfer excitations in nitrobenzene and N-phenylpyrrole, involving calculations of 4\textsuperscript{th}- and 6\textsuperscript{th}-order saddle points, respectively. Finally, calculations of a diplatinum and silver complex are presented, illustrating the applicability of the method to excited state energy curves of large molecules.
\end{abstract}

\section{Introduction}
Calculations of photochemical processes require electronic structure methods that can describe a wide range of excitations and give the variation of the energy of the excited states as a function of the atomic coordinates. Linear-response time-dependent density functional theory\cite{Runge1984, Casida1995} in the adiabatic approximation (henceforth referred to as TDDFT) is a commonly used method for the calculation of excited electronic states, due to its relatively low computational cost. However, TDDFT has some important limitations\cite{Herbert2022}. Firstly, due to the linear-response formalism, TDDFT with local and semi-local functionals cannot adequately describe excitations involving large changes in the electron density, such as charge transfer, core, and Rydberg excitations.\cite{Dreuw2004, Dreuw2005, Hait2021}. Secondly, as the atomic configuration changes, the reference state of a TDDFT calculation can abruptly change character, leading to a discontinuity in the energy of the excited state, thereby making the evaluation of atomic forces ill-defined. This behavior is, e.g., observed in bond stretching\cite{Hait2019, Hait2019_2} and in the vicinity of electronic degeneracy,\cite{Barca2018} such as a conical intersection. The latter can play a major role in photochemistry, as the Born-Oppenheimer approximation breaks down, and population transfer can readily occur between the electronic states. The topology of conical intersections is described incorrectly in TDDFT\cite{Levine2006} because doubly excited configurations are missing in the response state, and the degeneracy is thereby not lifted along one of the branching space atomic displacements. Several modifications of TDDFT have been proposed to improve the description of excited states, such as a configuration interaction using one doubly excited configuration,\cite{Athavale2021, Teh2019} the dual-functional approach,\cite{shu2017dual, Shu2017} the spin-flip TDDFT\cite{Shao2003}, and the particle-particle\cite{Yang2016} and hole-hole\cite{Bannwarth2020} Tamm-Dancoff approximations. The description of charge transfer states can be improved by using higher-level functional approximations, such as double hybrid\cite{Ottochian2020, Bremond2021} and optimally tuned\cite{Stein2009, Kronik2012, Korzdorfer2014} functionals, but these enhancements involve a significant increase in computational effort. Moreover, optimal tuning is system- and geometry-specific and is, therefore, of limited applicability in simulations of atomic dynamics in excited states.

Variational density functional calculations of excited states\cite{Vandaele2022, Hait2021, Levi2020, Carter-Fenk2020, Ayers2015} are emerging as an attractive alternative to TDDFT. They typically involve similar computational cost as ground state calculations and can better describe long-range charge transfer\cite{Hait2021,Barca2018,Zhekova2014}, Rydberg\cite{Seidu2015,Cheng2008}, core-level\cite{Besley2021,Besley2009}, and other excitations\cite{Hait2021,Hait2020} where a significant change in the electron density occurs. As the calculations are variational, they provide atomic forces that can be used in excited state geometry optimization and classical dynamics simulations\cite{Vandaele2022_2, Malis2020, Pradhan2018, Levi2018}. In a variational calculation within a mean-field approximation, an excited state is found as a solution of higher energy than the ground state and corresponds to an optimal single Slater determinant with non-aufbau orbital occupation. An important feature of an excited state stationary solution is that it typically corresponds to a saddle point on the surface that describes the variation of the energy as a function of the electronic degrees of freedom, while the ground state corresponds to a minimum.

On the exact electronic energy landscape, provided by full configuration interaction, the $n$\textsuperscript{th} excited state above the ground state corresponds to a saddle point of order $n$\cite{Burton2022, MolecularElectronicStructureTheory}. Accordingly, state-specific multiconfigurational self-consistent field (MCSCF) calculations of excited states have employed optimization strategies that specifically search for saddle points\cite{Jensen1984, Olsen1983, Golab1983} on the MCSCF energy surface. For mean-field approximations, excited states do not always correspond to saddle points, but it is usually observed that the number of negative eigenvalues of the electronic Hessian at a stationary solution increases with the energy (although not necessarily monotonically \cite{Burton2022}), so the mean-field approximate excited states are typically saddle points\cite{Hait2021, Levi2020, Hait2020, Perdew1985}. Commonly used methods for variational mean-field calculations of excited states have, however, not been specifically designed to find saddle points on the electronic energy surface. The most common approach is the SCF method based on Hamiltonian matrix diagonalization. Since the excitation energy is computed as the difference relative to the ground state energy, the term ``$\Delta$SCF'' is often used in the literature for such calculations \cite{Herbert2022, Vandaele2022, Vandaele2022_2, Malis2020, Kowalczyk2011, Pradhan2018, Levi2018}. SCF calculations of excited states typically use iterative eigensolvers developed for ground state calculations, such as the Davidson\cite{Davidson1975} method or the direct inversion in the iterative subspace (DIIS)\cite{Pulay1980, Pulay1982}. Additionally, a maximum overlap method\cite{Taka2022, Gilbert2008, Barca2018} (MOM) is often used to reduce the likelihood of converging on the ground state, i.e.\ variational collapse (we will refer to this approach of using an SCF procedure with MOM as SCF-MOM). Although methods based on SCF can converge on saddle points, they are better suited for the minimization needed for ground state calculations. Moreover, they can have convergence problems close to electronic near-degeneracies, e.g.\ in the vicinity of conical intersections or points at which symmetry-broken solutions arise, as are often encountered in bond breaking processes. 

Recently, a direct optimization (DO) method has been developed\cite{Ivanov2021,Ivanov2021_2,Levi2020} based on an exponential transformation and a quasi-Newton algorithm for finding the pairwise orbital rotations that make the energy stationary. This approach can converge on saddle points on the electronic energy surface when a preconditioned quasi-Newton method is used together with MOM (referred to as DO-MOM method) to maximize the energy along some electronic degree(s) of freedom corresponding to directions of negative curvature. Combined with the robustness typical of approximate second-order optimization in the vicinity of an electronic degeneracy\cite{Levi2020,Levi2020_2,VanVoorhis2002}, these attributes make DO-MOM a useful tool for variational excited state calculations. For example, by using DO-MOM, it has been shown\cite{Schmerwitz2022} that symmetry-broken ground and excited state solutions provide a qualitatively correct conical intersection and avoided crossing in the ethylene molecule, even when a semi-local generalized gradient approximation functional is used.\cite{Perdew1996, Perdew1997} The application of explicit self-interaction correction improves the calculated results and gives nearly quantitative agreement with high-level quantum chemistry calculations.\cite{Schmerwitz2022} A previous study using an SCF-MOM approach turned out to be less successful.\cite{Pradhan2018}

Despite these advancements in the algorithms for variational density functional calculations of excited states, some limitations remain which hamper widespread application of the methodology. One issue is related to the fact that multiple solutions providing qualitatively different variation of the energy as a function of the atomic coordinates (the molecular potential energy surface) can exist for a given excited state. For example, when photoexcitation involves bond breaking, different electronic configurations approach degeneracy, leading to increased static correlation and the emergence of multiple mean-field solutions, some conserving and others breaking the symmetries of the Hamiltonian.\cite{Jake2018,Toth2016,Jimenez-Hoyos2011,Li2009,Coulson1949} We will refer to atomic configurations where multiple mean-field solutions start to appear as symmetry-breaking onsets (SBOs). Typically, symmetry-broken solutions provide an adequate description of the energy surface beyond SBOs, due to their implicit multiconfigurational character\cite{Schmerwitz2022, Perdew2021, Yu2016, Cohen2008, Grafenstein2002, Cremer2002, Cremer2001, Grafenstein2000, Wittbrodt1996}. Other solutions, however, give qualitatively incorrect potential energy surfaces for atom dynamics, for instance corresponding to diabatic surfaces\cite{Schmerwitz2022}, which can lead to unphysical state switching in geometry optimizations and classical dynamics simulations. As the ultimate goal of variational excited state calculations is often the simulation of atomic dynamics in the excited state \cite{Malis2022, Vandaele2022_2, Malis2020, Pradhan2018}, it is important to ensure that the calculations converge on the solution providing the appropriate value of the energy and atomic forces. While the ground state solution can, by the minimum energy principle, be identified as the one with the lowest energy,\cite{Vaucher2017} an analogous method cannot be used for tracking excited state solutions.\cite{Schmerwitz2022, MolecularElectronicStructureTheory} In previous work\cite{Schmerwitz2022}, we demonstrated that the mean-field solution for the doubly excited state of ethylene giving the best description of the energy curve along the double bond torsion can be identified based on the conservation of the saddle point order as the wave function undergoes symmetry breaking. This approach provides a guiding principle for targeting a specific solution during the exploration of an excited state potential energy surface for atomic dynamics. A second remaining issue is problematic convergence on a high-order saddle point, especially when wave function relaxation leads to large rearrangements of the orbitals, as can be the case for charge transfer excitations. Although the DO-MOM approach offers an improvement over SCF-based methods\cite{Ivanov2021, Levi2020, Levi2020_2}, variational collapse can still be an issue, as will be demonstrated below. 

These challenges call for the development of a practical and robust method specifically designed to find a saddle point of a given order on the electronic energy surface. In the context of finding the mechanism and rate of atomic rearrangements, such as chemical reactions and diffusion events, several methods have been developed for finding 1\textsuperscript{st}-order saddle points on energy surfaces that describe the variation of the ground electronic state energy as a function of atomic coordinates.\cite{Pelzer1932, Eyring1935, Wigner1938} Convergence on a 1\textsuperscript{st}-order saddle point can be achieved by taking a step uphill in energy in the direction of the eigenvector corresponding to the lowest eigenvalue of the Hessian matrix and downhill in all perpendicular directions. This approach can be generalized to find higher-order saddle points by taking an uphill step along more than one eigenvector. If the Newton-Raphson algorithm is used, second-order local convergence can be achieved \cite{Cerjan1981, Simons1983, Banerjee1985}. Such algorithms have been developed for state-specific MCSCF calculations of excited states, typically involving the trust-region method for the Newton-Raphson step \cite{Marie2013, Hoffmann2002, Jensen1984, Golab1983}. This method is based on the evaluation of second derivatives of the energy, thereby involving large computational effort making it impractical for large systems. The step control with trust radius for excited states can, furthermore, be problematic\cite{Jensen1984}. Quasi-Newton methods use approximate Hessian update techniques, such as the symmetric rank-one (SR1),\cite{Murtagh1970} Powell,\cite{Powell1973} and Bofill\cite{Bofill1994} updates, and require less computational effort. This approach has been used in DO-MOM calculations of excited states, \cite{Ivanov2021, Levi2020, Levi2020_2} but it relies on having a good initial guess for the Hessian to construct a preconditioner. 

An alternative approach for finding 1\textsuperscript{st}-order saddle points that does not involve constructing or estimating the Hessian and still provides fast convergence is the minimum mode following method \cite{Henkelman1999,Olsen2004,Kastner2008,Gutierrez2016}. There, only the eigenvector of the Hessian corresponding to the lowest eigenvalue (referred to as the minimum mode) needs to be evaluated, which can be accomplished, for example, by using the dimer \cite{Henkelman1999}, L{\'a}nczos \cite{Lanczos1950,Olsen2004} or Davidson \cite{Davidson1975,Gutierrez2016} methods. The projection of the gradient on the direction of the minimum mode is then inverted and the optimization performed using any gradient-based minimization technique.

Here, the minimum mode following method is generalized to target an $n$\textsuperscript{th}-order saddle point by following the eigenvectors of the $n$ lowest eigenvalues instead of just the minimum mode. The needed eigenvectors are evaluated using a generalized Davidson algorithm\cite{Crouzeix1994}, thereby requiring only first-order derivatives of the energy. The resulting generalized mode following (GMF) approach is combined with the exponential transformation direct optimization of the orbitals to yield the DO-GMF method. It can be used to calculate excited states corresponding to saddle points without the need of MOM or other special procedures to reduce the risk of variational collapse. To illustrate the methodology and demonstrate its performance, DO-GMF calculations of excited states with the PBE generalized gradient approximation functional\cite{Perdew1996} are presented here. The convergence properties are compared to those of the DO-MOM method, which has previously been shown to outperform SCF-MOM-based strategies\cite{Ivanov2021, Levi2020, Levi2020_2}. First, the energy curves of the doubly excited states of the dihydrogen and ethylene molecules along the single bond stretching and double bond torsion, respectively, are calculated. These calculations are performed with DO-GMF by targeting a 2\textsuperscript{nd}-order saddle point, while DO-MOM converges on a lower-energy solution after the SBO, giving a qualitatively incorrect description of the potential energy surface for atomic dynamics. The performance of DO-GMF is further assessed by calculating charge transfer excited states of the nitrobenzene and twisted N-phenylpyrrole molecules as well as the bond dissociation energy curve of the first excited state of a large diplatinum and silver complex, [AgPt$_2$(P$_2$O$_5$H$_2$)$_4$]$^{3-}$ (AgPtPOP). The DO-GMF approach is found to be more robust than the DO-MOM method in calculations of excited states using density functionals and furthermore, ensures convergence to a solution providing potential energy curves consistent with the target excited state for atomic dynamics at SBOs.

The present article is structured as follows. In section 2, the methodology is presented, including a summary of direct orbital optimization and the saddle point search algorithm. The correspondence between excited states and saddle points on the electronic energy surface is illustrated for the minimal-basis H$_2$ molecule. Section 3 gives information on the implementation and parameters used in the density functional calculations. Results are presented in section 4, where the ability of DO-GMF to track excited state solutions through SBOs is demonstrated by calculations of potential energy curves for the dihydrogen and ethylene molecules. Additionally, the performance of DO-GMF and DO-MOM is compared in calculations of charge transfer excitations of nitrobenzene and twisted N-phenylpyrrole as well as in calculations of the energy curve of the first excited state of AgPtPOP. The article concludes with a discussion of the results and a future outlook in section 5. 

\section{Methodology}
\subsection{Direct Orbital Optimization}
DO\cite{Levi2020, Lehtola2020} is an alternative to the SCF approach. In DO, one seeks a unitary matrix $\matr{U}$ transforming an initial guess (or reference) electronic wave function $\vect{\psi}_{0}$ into the wave function $\vect{\psi}_{\mathrm{stat}}$ that represents a stationary point on the electronic energy surface. Within the linear combination of atomic orbitals (LCAO) formalism, the molecular orbitals, $\vect{\psi} = \left(\ket{\psi_1}, \ldots, \ket{\psi_{M}}\right)^{T}$\,, are expanded in a set of $M$ basis functions, $\vect{\phi} = \left(\ket{\phi_1}, \ldots, \ket{\phi_{M}}\right)^{T}$\,, as
\begin{equation}
\vect{\psi} = \matr{C}\vect{\phi}\,,
\end{equation}
where $\matr{C}$ is the $M \times M$ matrix of coefficients. Thus, the objective is to find the optimal orbital coefficients $\matr{C}_{\mathrm{stat}}$ that make the energy stationary by applying a unitary transformation to a set of reference orbitals $\matr{C}_{0}$\,,
\begin{equation}
    \label{eq:unitary-transformation}
    \textbf{C}_{\mathrm{stat}} = \matr{C}_{0}\matr{U}\,.
\end{equation}
In the Hartree-Fock and Kohn-Sham (KS) approaches, the ground state is the global minimum of the energy as a function of $\matr{U}$, while excited electronic states are typically saddle points.\cite{Burton2022, Hait2021, Levi2020, Perdew1985}

\subsubsection{Exponential Transformation}
The unitary matrix can conveniently be expressed as the exponential of a matrix $\matr{\Theta}$ containing pairwise orbital rotation angles. The orbital orthonormality constraints are imposed by requiring $\matr{\Theta}$ to be anti-Hermitian,
\begin{equation}
    \matr{U} = e^{\matr{\Theta}}, \mathrm{\ subject\ to\ } \matr{\Theta} = -\matr{\Theta}^{\dag}\,.
\end{equation}
A stationary point in the linear space of anti-Hermitian matrices directly corresponds to a stationary point in the non-linear space of the corresponding exponentials. Therefore, the problem is recast into making the energy stationary as a function of the elements of $\matr{\Theta}$, which can be decomposed into three unique blocks containing rotations among the occupied orbitals (oo), rotations among the virtual orbitals (vv), and rotations between the occupied and virtual orbitals (ov),
\begin{equation}
    \matr{\Theta} = \begin{pmatrix}\matr{\Theta}_{\mathrm{oo}} & \matr{\Theta}_{\mathrm{ov}}\\ -\matr{\Theta}_{\mathrm{ov}}^{\dag} & \matr{\Theta}_{\mathrm{vv}}\end{pmatrix}\,.
\end{equation}
The energy gradient with respect to rotations in the vv block is always 0, meaning that this block is constant during an optimization and can without loss of generality be set to 0. The oo block can be set to 0 if the density functional is unitary invariant. However, if orbital density dependence is incorporated into the functional, e.g.\ by including the Perdew-Zunger self-interaction correction\cite{Perdew1981}, the functional is no longer unitary invariant, meaning that the optimal $\matr{\Theta}_{\text{oo}}$ is a unique anti-Hermitian matrix.\cite{Lehtola2016} The ov block always has to be optimized. The size of the oo, ov and vv blocks is $N\times N$, $N\times (M-N)$ and $(M-N)\times (M-N)$, respectively, $N$ being the number of occupied orbitals, though only one triangular part of the oo block is unique for orbital density dependent functionals.

The optimization is performed in the linear space of anti-Hermitian matrices by applying standard numerical optimization techniques, such as quasi-Newton methods. The optimization method used for excited state calculations must be able to converge on saddle points. The components of the gradient are
\begin{equation}
    \label{eq:gradient}
    \frac{\partial E}{\partial \Theta_{ij}} = \frac{2 - \delta_{ij}}{2}\left[\int_{0}^{1}dte^{t\matr{\Theta}}\matr{L}e^{-t\matr{\Theta}}\right]_{ji} = \frac{2 - \delta_{ij}}{2}\left[\matr{L} + \frac{1}{2!}\left[\matr{\Theta}, \matr{L}\right] + \frac{1}{3!}\left[\matr{\Theta}, \left[\matr{\Theta}, \matr{L}\right]\right] + ...\right]_{ji}
\end{equation}
with
\begin{equation}
    L_{ij} = \left(f_{i} - f_{j}\right)H_{ij}\,,
\end{equation}
where $H_{ij}$ are the elements of the Hamiltonian matrix in the basis of molecular orbitals and $f_i$ are the orbital occupation numbers. The right-hand side of eq\,\ref{eq:gradient} is a special case of the Baker-Campbell-Hausdorff formula\cite{Baker1905, Campbell1896, Campbell1897, Hausdorff1906} and makes use of the fact that the anti-Hermitian matrices form the Lie-algebra corresponding to the Lie-group of unitary matrices. If the norm of $\matr{\Theta}$ is small (${\| \matr{\Theta} \|}  \ll 1$), the series can be truncated at the first term ($\frac{\partial E}{\partial \Theta_{ij}} \approx \frac{2 - \delta_{ij}}{2}L_{ji}$), providing an efficient way of evaluating the gradient. The elements of $\matr{\Theta}$ can be kept small by updating the reference orbitals and setting all orbital rotations to 0 at regular step intervals. For a unitary invariant functional, the following diagonal approximation to the Hessian can be used as preconditioner,
\begin{equation}
    \label{eq:diagonal_hessian}
    \frac{\partial^{2}E}{\partial \Theta_{ij}^{2}} \approx 2\left(f_{j} - f_{i}\right)\left(\epsilon_{i} - \epsilon_{j}\right)\,,
\end{equation}
where $f_{i}$ and $\epsilon_{i}$ are the orbital occupation numbers and eigenvalues of the Hamiltonian matrix, respectively. 

An excited state calculation is usually initialized using the ground state orbitals after swapping the occupation numbers of the occupied and unoccupied orbitals involved in the excitation. For example, in a HOMO-LUMO excitation, the HOMO gets assigned an occupation number of 0 while the LUMO gets assigned an occupation number of 1. Eq \ref{eq:diagonal_hessian} is then used with the resulting non-aufbau occupations and the energies of the ground state orbitals. The preconditioner in an excited state calculation has one negative eigenvalue for each pair of occupied-unoccupied orbitals where the energy of the unoccupied orbital is smaller than the energy of the occupied orbital.

\subsection{Excited states and saddle points on the energy surface}
\begin{figure}
    \centering
    \includegraphics[width=0.65\textwidth]{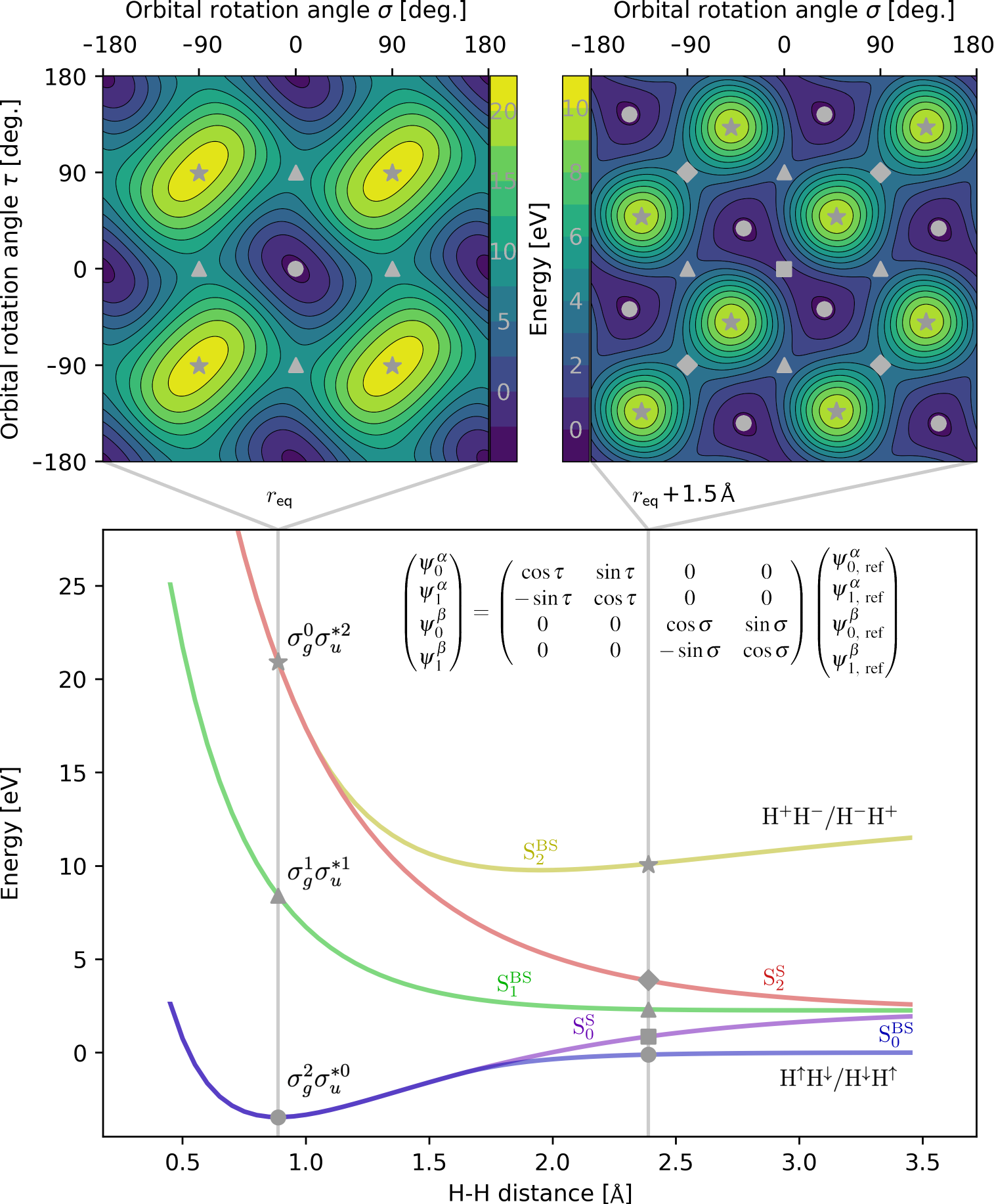}
    \caption{
        Energy of the three states with one electron in each spin channel, $\alpha$ and $\beta$, that can be obtained for the minimal-basis H$_{2}$ molecule as a function of the atom distance. The orbitals are related by rotation angles $\sigma$ and $\tau$ with respect to the orbitals of the symmetry-pure ground state solution, S$_{0}^{\mathrm{S}}$, with configuration $\sigma_g^2\sigma_u^{*0}$: $\psi^{\alpha}_{0,\mathrm{ref}}=\psi^{\beta}_{0,\mathrm{ref}}=\sigma_{g}$ and  $\psi^{\alpha}_{1,\mathrm{ref}}=\psi^{\beta}_{1,\mathrm{ref}}=\sigma_{u}^{*}$, where $\sigma_{g}$ and $\sigma_{u}^{*}$ are symmetry-adapted bonding and antibonding molecular orbitals, respectively. The subscripts $0$ and $1$ indicate occupied and unoccupied orbitals, respectively. The ground state corresponds to a minimum on the electronic energy surface shown for the optimal bond length, $r_{e}$, in the left contour graph, while the first and doubly excited state solutions, S$_{1}^{\mathrm{BS}}$ and S$_{2}^{\mathrm{S}}$, correspond to 1\textsuperscript{st}- and 2\textsuperscript{nd}-order saddle points, respectively. Beyond a distance of 1.1\,\AA, the spatial symmetry  can break in the second excited state to give a higher-energy solution, S$_{2}^{\mathrm{BS}}$, corresponding to charge transfer between the two H atoms. Beyond a distance of 1.7\,\AA, the spin symmetry can break in the ground state to give a lower-energy solution, S$_{0}^{\text{BS}}$, corresponding to spin polarization between the two H atoms. The right contour graph shows the energy surface at $r_{e} + 1.5$\,\AA , which is beyond the two symmetry breaking points. There, the extrema have each split into two, with the minima and maxima corresponding to symmetry-broken ground and doubly excited state solutions, S$_{0}^{\mathrm{BS}}$ and S$_{2}^{\mathrm{BS}}$, respectively, while the symmetry-pure solutions, S$_{0}^{\mathrm{S}}$ and S$_{2}^{\mathrm{S}}$, correspond to 1\textsuperscript{st}-order saddle points.}
    \label{fig:main_h2_pes_plot}
\end{figure}

The correspondence between saddle points on the electronic energy surface and excited electronic states is illustrated in Figure\ \ref{fig:main_h2_pes_plot} using the H\textsubscript{2} molecule with a minimal basis set as an example. The calculations are carried out with the PBE functional\cite{Perdew1996} in the spin-unrestricted formalism. 
There are only two molecular orbitals for each spin channel in this case for the two electrons. The figure shows the potential energy curves for all solutions that can be obtained with one electron in each spin channel. Contour graphs of the electronic energy surface are also shown to illustrate how the energy depends on the rotation angles $\tau$ and $\sigma$ that mix the occupied and unoccupied orbitals in each spin channel. One contour graph is shown for the equilibrium bond length and another for a stretched bond where symmetry has been broken in both the ground state and the doubly excited state. Each point on this electronic energy surface is obtained according to the transformation
\begin{equation*}
    \begin{pmatrix}
        \vspace{1pt}
        \psi_{0}^{\alpha}\\
        \vspace{1pt}
        \psi_{1}^{\alpha}\\
        \vspace{1pt}
        \psi_{0}^{\beta}\\
        \vspace{1pt}
        \psi_{1}^{\beta}
    \end{pmatrix}
    =
    \begin{pmatrix}
            \cos{\tau} & \sin{\tau} & 0 & 0\\
            -\sin{\tau} & \cos{\tau} & 0 & 0\\
            0 & 0 & \cos{\sigma} & \sin{\sigma}\\
            0 & 0 & -\sin{\sigma} & \cos{\sigma} 
    \end{pmatrix}
    \begin{pmatrix}
        \vspace{1pt}
        \psi_{0,\,\mathrm{ref}}^{\alpha}\\
        \vspace{1pt}
        \psi_{1,\,\mathrm{ref}}^{\alpha}\\
        \vspace{1pt}
        \psi_{0,\,\mathrm{ref}}^{\beta}\\
        \vspace{1pt}
        \psi_{1,\,\mathrm{ref}}^{\beta}
    \end{pmatrix}\,,
\end{equation*}
where the indices 0 and 1 correspond to occupied and unoccupied orbitals, respectively, and the indices $\alpha$ and $\beta$ correspond to the two spin channels. The reference orbitals (subscript ``ref'') are the orbitals of the symmetry-pure ground state solution, S$_{0}^{\text{S}}$, with configuration $\sigma_g^2\sigma_u^{*0}$, where $\sigma_g$ and $\sigma_u^{*}$ are the symmetry-adapted bonding and antibonding molecular orbitals, respectively. At the ground state equilibrium geometry, S$_{0}^{\text{S}}$ corresponds to the global minimum on the electronic energy surface. There are two other types of solutions at this geometry: a solution with broken spin symmetry, S$_{1}^{\text{BS}}$, representing a singly excited open-shell state with configuration $\sigma_g^1\sigma_u^{*1}$ corresponding to a 1\textsuperscript{st}-order saddle point obtained by $\pm 90^{\circ}$ rotation in one of the spin channels, and a symmetry-pure solution representing a doubly excited state with configuration $\sigma_g^0\sigma_u^{*2}$ corresponding to a 2\textsuperscript{nd}-order saddle point obtained by $\pm 90^{\circ}$ rotation in both spin channels. For each symmetry-pure state, there are two degenerate solutions related by a sign change, while there are four degenerate solutions for each symmetry-broken state additionally related by a symmetry operation. 

As the bond is stretched, SBOs appear from which two additional types of solutions emerge. One is a solution with broken spin symmetry, S$_{0}^{\text{BS}}$, representing a ground state with diradical character (H$^{\uparrow}$H$^{\downarrow}$/H$^{\downarrow}$H$^{\uparrow}$). The SBO for this symmetry-broken ground state solution is often referred to as a Coulson-Fischer point\cite{Coulson1949}. S$_{0}^{\text{BS}}$ corresponds to the global minimum on the electronic energy surface, and as the H-H distance tends to infinity, it can be obtained from the symmetry-pure S$_{0}^{\text{S}}$ solution by applying $45^{\circ}$ orbital rotations of opposite sign in the two spin channels. The other symmetry-broken solution, S$_{2}^{\text{BS}}$, represents a doubly excited state with ionic character (H$^+$H$^-$/H$^-$H$^+$) where the spatial symmetry is lost. It is a 2\textsuperscript{nd}-order saddle point accessible through $45^{\circ}$ orbital rotations of the same sign in both spin channels as the H-H distance tends to infinity. The symmetry-broken solutions, S$_{0}^{\text{BS}}$ and S$_{2}^{\text{BS}}$,  arise as degenerate pairs related by a swap of the spin channels and inversion at the center of the molecule, respectively. They thereby cease to be eigenfunctions of those symmetry operations. 
 
The symmetry-broken states are the preferred solutions in a mean-field calculation as they effectively take static correlation into account\cite{Perdew2021, Cremer2001}. In a calculation of an energy curve representing the potential energy profile for atomic dynamics, it is important to identify which solution to follow at SBOs. As can be seen from Figure\ \ref{fig:main_h2_pes_plot}, the saddle point order for the symmetry-broken solutions is the same as that of the symmetry-pure counterpart before the SBOs. Following a saddle point of a given order, therefore, smoothly tracks the symmetry-broken solution through an SBO. The symmetry-pure S$_{0}^{\text{S}}$ and S$_{2}^{\text{S}}$ solutions, however, become 1\textsuperscript{st}-order saddle points beyond the SBOs. Therefore, the SBOs represent inflection points on the energy curves of the symmetry-pure solutions (a sign change of the first and second Hessian eigenvalues for S$_{0}^{\text{S}}$ and S$_{2}^{\text{S}}$, respectively), while no inflection points are present for the symmetry-broken solutions (persistent sign of the first/second Hessian eigenvalue). The symmetry-broken solutions provide potential energy curves for the bond dissociation with qualitatively correct shape and asymptotic behavior\cite{Salem1972}, while the symmetry-pure solutions overestimate the energy of the ground state and underestimate the energy of the doubly excited state, the two becoming degenerate at infinite H-H distance.

\subsubsection{Computing excited state energy surfaces}
There are two ways of mapping out excited state potential energy surfaces for the dynamics of atoms in molecules, such as those illustrated in Figure\,\ref{fig:main_h2_pes_plot}. The strategy that is most commonly employed in geometry optimizations and classical dynamics simulations is a sequential point acquisition scheme. Here, one first performs a ground state calculation at the initial geometry and uses the obtained orbitals with occupation numbers changed to reflect the character of the target excited state as an initial guess for the excited state wave function optimization. Subsequent points are generated by displacing the atoms and using the orbitals of the previous converged solution as the initial guess. The second strategy is a separate point acquisition scheme, where for each configuration of the atoms the ground state orbitals are used as the initial guess for the excited state wave function calculation. The sequential scheme is more efficient, as it does not require a ground state calculation for each geometry and, moreover, can converge faster if the atomic displacements are sufficiently small. However, as mentioned above, a problem can arise when multiple solutions emerge from one at an SBO, if the calculation converges to a solution that gives a qualitatively incorrect potential energy surface. As illustrated by calculations of the excited state of H$_2$ discussed above and also by our previously published calculations of the torsional potential energy curves for the ethylene molecule\cite{Schmerwitz2022}, the multiple excited state solutions correspond to saddle points of different orders. For excited electronic states, one cannot simply select the state that has lower energy, as for the ground state, but a possibility is to select the state that corresponds to the same saddle point order before and after the SBO. The mapping of excited state potential energy surfaces, therefore, requires a method for reconverging on a saddle point of a given order after the atomic configuration has been changed. Below we present such a method, the generalized mode following approach.

\subsection{Generalized Mode Following}
\subsubsection{Minimum Mode Following}
Minimum mode following is a saddle point optimization technique used to determine 1\textsuperscript{st}-order saddle points on the energy surface in atomic configuration space \cite{Henkelman1999,Olsen2004,Gutierrez2016}. The method recasts the challenging saddle point search as a minimization by inverting the projection of the gradient vector $\vect{g}$ on the unstable Hessian mode $\vect{v}_{1}$ at each iteration of the optimization,
\begin{align}
    \label{eq:mmf_1}
    \vect{g}^{\,\parallel} & = \vect{v}_{1}\vect{v}_{1}^{\mathrm{T}}\vect{g}\,,\\ 
    \label{eq:mmf_2}
    \vect{g}^{\mathrm{\,mod}} & = \vect{g} - 2\vect{g}^{\,\parallel} & \mathrm{\ if\ } \lambda_{1} < 0\,,\\ 
    \label{eq:mmf_3}
    \vect{g}^{\mathrm{\,mod}} & = -\vect{g}^{\,\parallel} & \mathrm{\ if\ } \lambda_{1} \geq 0\,.
\end{align}
The perpendicular component of the force is not included (eq \ref{eq:mmf_3}) if the curvature $\lambda_{1}$ along the lowest eigenvector is positive, i.e.\ if the function is convex in this direction. This procedure yields a gradient that leads to a minimum in a revised objective function where the original function has a 1\textsuperscript{st}-order saddle point, so any gradient-based minimization technique can then be used to converge on the saddle point. However, since the modified objective function is not known, line search techniques commonly used to accelerate convergence cannot straightforwardly be applied. 

The lowest curvature Hessian mode can be obtained with a variety of partial diagonalization methods. The most common ones are the dimer\cite{Henkelman1999, Kastner2008, Gould2014}, Davidson\cite{Davidson1975, Crouzeix1994} and Lánczos\cite{Lanczos1950} methods, where the relevant parts of the Hessian are evaluated by a finite difference approximation. The Davidson method can be regarded as a preconditioned Lánczos method. The preconditioner is typically diagonal but a full approximate Hessian can also be used.

\subsubsection{Generalization to saddle points of arbitrary order}
While the estimation of rates of atomic rearrangements involves finding 1\textsuperscript{st}-order saddle points, the calculation of excited electronic states can require finding saddle points of higher order. The mode following method can be generalized to find a saddle point of order $n$ by identifying the modes corresponding to the $n$ lowest eigenvalues and following them simultaneously. The gradient projections along the $n$ modes are summed up to yield the parallel gradient which is then inverted in the usual way,
\begin{align}
    \label{eq:gmf_1}
    \vect{g}^{\,\parallel} & = \sum_{i = 1}^{n}\vect{v}_{i}\vect{v}_{i}^{\mathrm{T}}\vect{g} \ \ \ {\mathrm{ and}} \ \ \ \vect{g}^{\mathrm{\,mod}}  = \vect{g} - 2\vect{g}^{\,\parallel} & \mathrm{\ if\ } \lambda_{n} < 0\,,\\
    \label{eq:gmf_2}
    \vect{g}^{\,\parallel} & = \sum_{\substack{i = 1 \\[4pt] \lambda_{i} \geq 0}}^{n}\vect{v}_{i}\vect{v}_{i}^{\mathrm{T}}\vect{g} \ \ \ {\mathrm { and}} \ \ \ \vect{g}^{\mathrm{\,mod}}  = -\vect{g}^{\,\parallel} & \mathrm{\ if\ } \lambda_{n} \geq 0\,.
\end{align}
Note that the perpendicular component of the gradient is only included in eq. \ref{eq:gmf_1} if eigenvalue $n$ is negative, i.e.\ if the electronic energy surface is concave along all the target eigenvectors. If this condition is not satisfied, only the target eigenvectors corresponding to non-negative eigenvalues are followed (eq. \ref{eq:gmf_2}).

\subsubsection{Implementation of GMF}
The implementation makes use of the generalized Davidson method presented in ref~\citenum{Crouzeix1994} to obtain the lowest $n$ eigenvectors of the electronic Hessian when targeting a saddle point of order $n$. At the first step of the wave function optimization, a diagonal approximation $\matr{D}$ to the Hessian is computed according to eq \ref{eq:diagonal_hessian}. Then, the generalized Davidson algorithm starts by defining an initial matrix $\matr{K}$ having  $n$ unit vectors $\vect{k}_i$ as columns along the orbital rotations corresponding to the lowest $n$ eigenvalues of the approximate Hessian (a Krylov subspace). A small random perturbation is applied to this initial Krylov subspace and the subspace is orthonormalized using the modified Gram-Schmidt approach. At each iteration of the Davidson algorithm, the effect of the electronic Hessian matrix $\mathscrbf{H}$ on the Krylov subspace is evaluated by a forward finite difference approximation
\begin{equation}
\mathscrbf{H}\vect{k}_{j} \approx \frac{\nabla E\left(\matr{C}e^{h\matr{\Theta}\left[\vect{k}_{j}\right]}\right) - \nabla E\left(\matr{C}\right)}{h}\,,
\end{equation}
where $\matr{\Theta}\left[\vect{k}_{j}\right]$ is the anti-Hermitian matrix having the elements of the $j$\textsuperscript{th} vector of the Krylov subspace, $\vect{k}_{j}$, in the upper triangular part, $h$ is the finite difference step size and $\nabla E\left(\matr{C}\right)$ is the energy gradient vector $\vect{g}$ evaluated using the current LCAO coefficient matrix $\matr{C}$. Next, the Rayleigh matrix $\matr{K}^{\mathrm{T}}\mathscrbf{H}\matr{K}$ is diagonalized to obtain the lowest $n$ eigenpairs ($\lambda_i$, $\vect{y}_i$). The Ritz vectors $\vect{x}_i = \matr{K}\vect{y}_i$ represent approximations to the target eigenvectors. The residual vectors $\vect{r}_i=(\lambda_i\matr{I}-\mathscrbf{H})\vect{x}_i$ are multiplied by a preconditioner $\matr{P}_i$, and the Krylov subspace is extended by incorporating the resulting $n$ vectors $\matr{P}_i\vect{r}_i$.  The preconditioner presented in ref ~\citenum{Sharada2015} is used,
\begin{equation}
\matr{P}_i=(\lambda_i\matr{I}-\matr{D})^{-1}\,,
\end{equation}
where $\matr{D}$ is the diagonal approximation to the Hessian given by eq \ref{eq:diagonal_hessian}. A threshold of $-0.1$\,Ha is applied to the components of the preconditioner. If it is exceeded for a given element, the element is set to this threshold to ensure that the preconditioner is negative definite and that the Davidson method converges to the lowest eigenpairs. At a given iteration, only the effect of the Hessian on the at most $n$ vectors added to the Krylov subspace in the previous iteration needs to be evaluated. If the dimensionality of the Krylov subspace becomes too large, a new Krylov subspace is formed by including only the $n$ approximate eigenvectors $\vect{x}_i$ and preconditioned residual vectors so that the cost of the subspace eigendecomposition is always small. The residuals $\vect{r}_i$ are also used to check for convergence. In the present implementation, the convergence threshold is a maximum component of 0.01\,Ha. If a target eigenpair is converged in a given iteration, the Krylov subspace is not extended in the preconditioned direction of the residual. Once all eigenpairs have converged and a step in the wave function optimization has been taken using the modified gradient (eq \ref{eq:gmf_1} or \ref{eq:gmf_2}), the next Davidson cycle is accelerated by using the eigenvectors found at the previous wave function optimization step to form the initial Krylov subspace. 
\begin{figure}
    \centering
    \includegraphics[width = \textwidth]{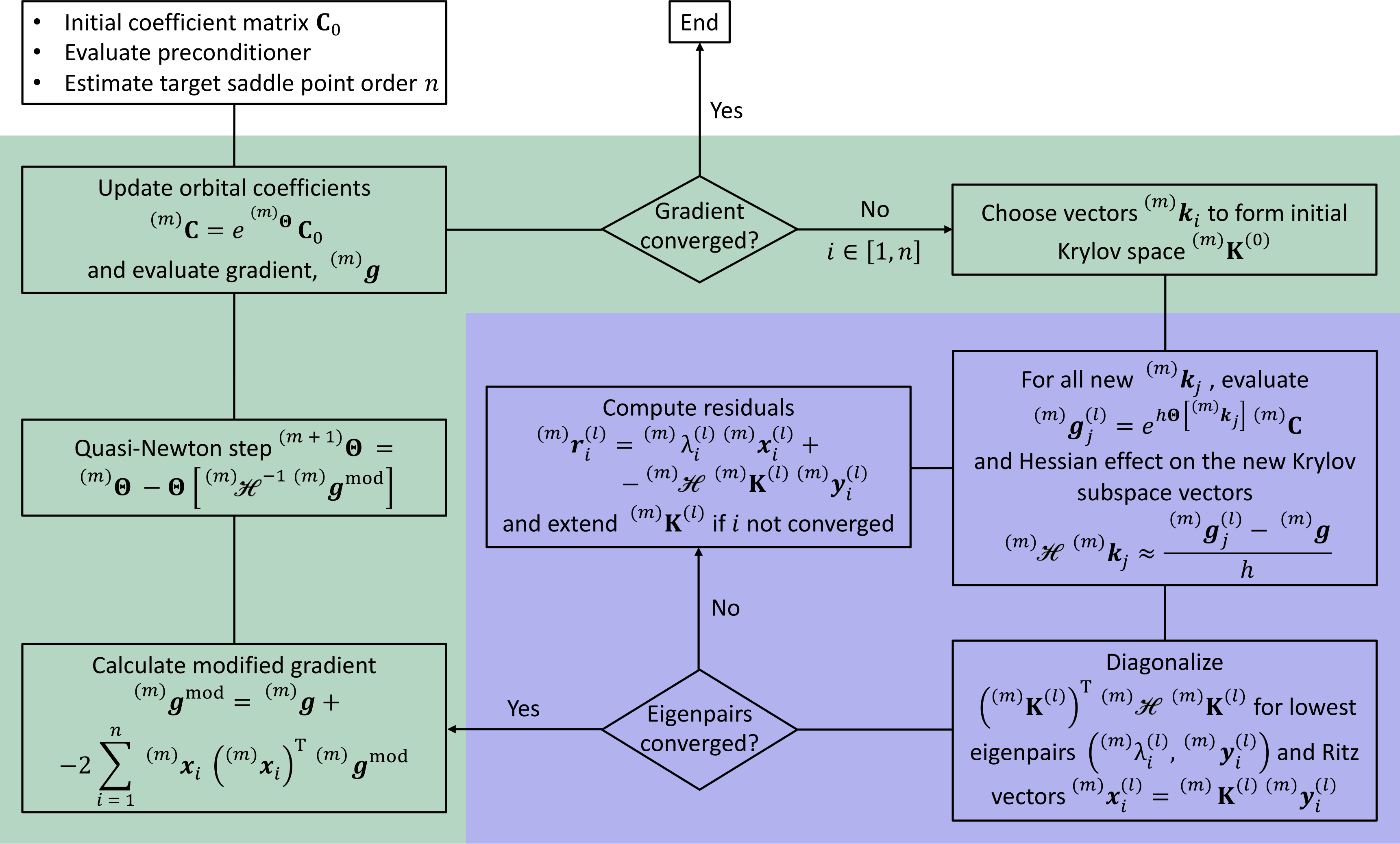}
    \caption{Flowchart of the DO-GMF algorithm. The approach consists of a direct optimization outer loop (green) using the exponential transformation and a quasi-Newton step with a modified gradient determined in a partial Hessian diagonalization inner loop (blue) using the generalized Davidson method.}
    \label{GMF_flowchart}
\end{figure}

As GMF recasts the saddle point search as a minimization, there is no risk of convergence on a lower-order saddle point or the ground state minimum (i.e.\ a variational collapse) and therefore, methods designed to reduce such problems, in particular MOM, are not needed. Additionally, there is no need for a quasi-Newton Hessian update with the flexibility to develop an indefinite matrix. Instead, one can use robust update formulas for minimization, such as the efficient limited-memory BFGS (L-BFGS) algorithm. The 
implementations of the exponential transformation and the quasi-Newton algorithms, including L-BFGS, for updating the orbitals in the LCAO representation are presented in refs ~\citenum{Levi2020} and ~\citenum{Ivanov2021_2}. An overview of the DO-GMF method is provided in the flowchart in Figure \ref{GMF_flowchart}.

The DO-GMF method has been implemented in a development branch of the grid-based projector augmented wave (GPAW)\cite{Mortensen2005, Enkovaara2010} software using the exponential transformation\cite{Levi2020} to update the orbitals represented in an LCAO basis\cite{Larsen2009}.

\subsection{Specifics of the electronic structure calculations}
The calculations make use of the generalized gradient approximation (GGA) functional PBE, apart from the calculations on the AgPtPOP complex, which instead use the BLYP functional. Valence electrons are represented by an LCAO basis set consisting of primitive Gaussian functions taken from the aug-cc-pVDZ\cite{Dunning1989, Kendall1992, Woon1994} (dihydrogen, ethylene), def2-TZVPD\cite{Weigend2005} (nitrobenzene), or cc-pVDZ (N-phenylpyrrole) sets augmented with a single set of numeric atomic orbitals\cite{Rossi2015, Larsen2009}. For the AgPtPOP complex, a double zeta polarized (dzp) basis set of numeric atomic orbitals\cite{Larsen2009} is used. Real orbitals are used. The frozen core approximation is used within the PAW approach\cite{Blochl1994}. 

The ground state calculations as well as the DO-MOM and DO-GMF excited state calculations are carried out with the exponential transformation direct optimization method \cite{Levi2020,Ivanov2021,Ivanov2021_2} using a limited-memory BFGS (L-BFGS) algorithm with inexact line search for the ground state\cite{Ivanov2021_2}, a limited-memory SR1 (L-SR1) algorithm with a maximum step length of 0.2 for the excited state DO-MOM calculations\cite{Levi2020} (default maximum step length in GPAW), and L-BFGS with a maximum step length of 0.2 for the excited state DO-GMF calculations, unless otherwise stated. The DO-MOM calculations use the MOM method as presented in ref ~\citenum{Barca2018}, where at each wave function optimization step, the occupied orbitals are chosen as those with the largest projections~\cite{Levi2020}
\begin{equation}
\omega_j = \sqrt{\sum_{i=1}^N |O_{ij}|^2}\,,
\end{equation}
where $O_{ij}$ are the elements of the overlap matrix between the orbitals at the current step and at the initial step
\begin{equation}
\matr{O} = \matr{C}_{0}^{\dag}\matr{S}\matr{C}
\end{equation}
with $\matr{S}$ being the overlap matrix of basis functions. No orthogonality constraints to lower-energy states are enforced, so the calculations are fully variational. The DO-MOM and DO-GMF calculations are converged to a precision of $10^{-7.4}\,\mathrm{eV}^{2}$ per valence electron in the squared residual of the KS equations and $10^{-7}$\,eV per valence electron in the three-step energy change. The atomic configuration of the nitrobenzene molecule is taken from ref \citenum{Hait2020}, while the N-phenylpyrrole molecule is from ref\,\citenum{Loos2021}. The atomic configuration of the ethylene molecule and the AgPtPOP complex are first optimized in the ground state at the same level of theory as used in the excited state calculations. The potential energy curves are then evaluated by incrementing a structural parameter while keeping the values of all other parameters the same as in the ground state optimal configuration. All open-shell singlet excited states are spin-mixed states, and spin purification is not applied to the energy. The calculations are performed with the GPAW software\cite{Mortensen2005, Enkovaara2010} and LIBXC\cite{Lehtola2018} version 4.0.4. The grid spacings are 0.14\,Å (dihydrogen with minimal basis set, AgPtPOP), 0.2\,Å (dihydrogen, ethylene) and 0.15\,Å (nitrobenzene, N-phenylpyrrole), while the dimensions of the simulation cell are according to the default cutoff of the numeric representation of the basis functions\cite{Rossi2015}. 

\subsection{Advantage of DO-GMF for finding symmetry-broken solutions}
\begin{figure}[!]
    \centering
    \includegraphics[width=\textwidth]{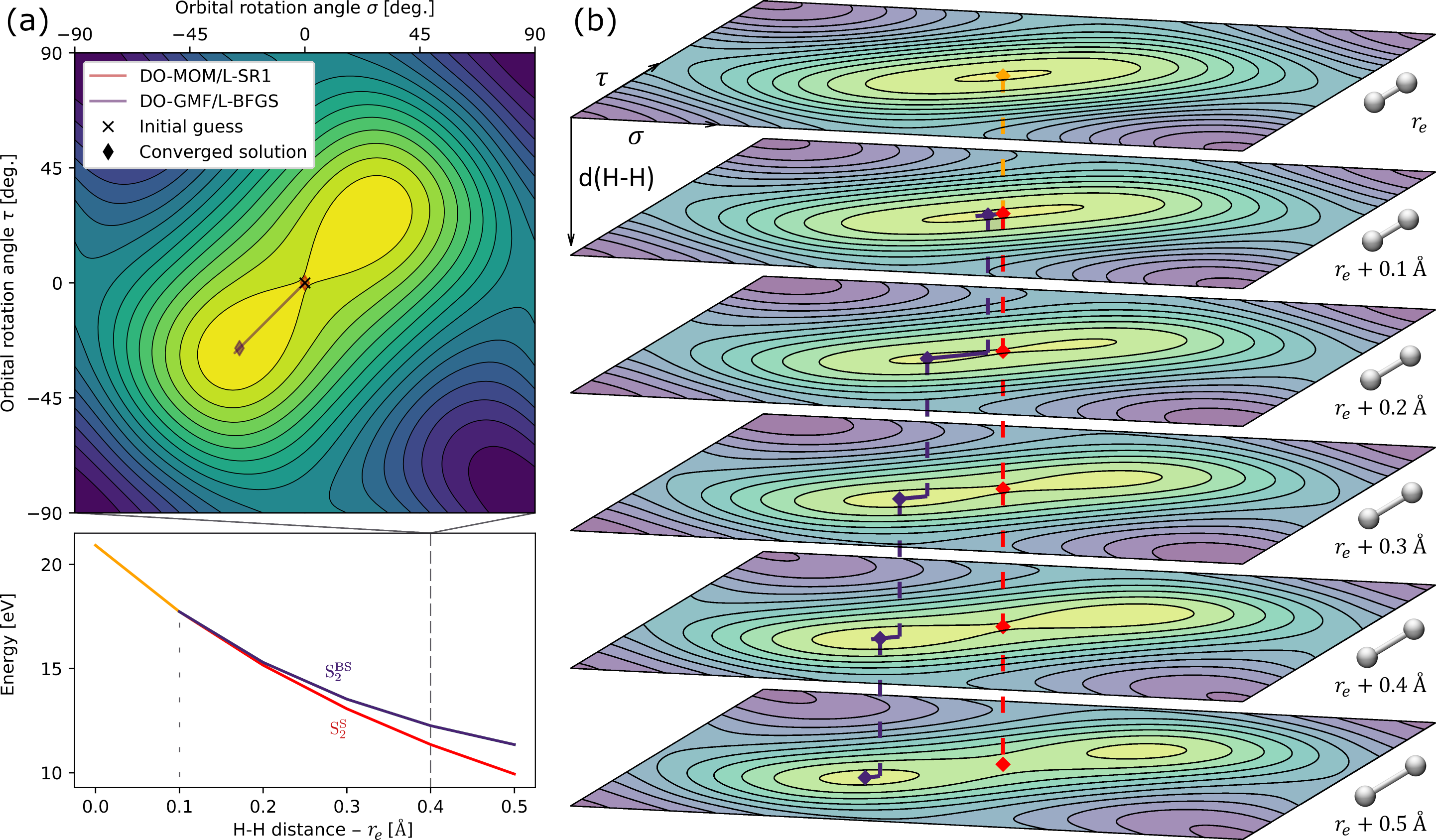}
    \caption{Illustration of the way DO-GMF calculations can converge on symmetry-broken solutions and provide a potential energy curve for atomic dynamics simulations, while DO-MOM converges on the symmetry-pure solutions. Results of calculations on the minimal-basis H\textsubscript{2} molecule in the doubly excited singlet state are shown using the PBE functional. (a) Single point calculations with DO-GMF and DO-MOM at $r_{e} + 0.4$\,\AA\ where symmetry breaking can occur. The initial guess (cross) is generated by double excitation from the symmetry-pure ground state solution, S$_{0}^{\text{S}}$. The DO-MOM calculation (red line) converges to the 1\textsuperscript{st}-order saddle point (red diamond) corresponding to the symmetry-pure solution, S$_{2}^{\text{S}}$, as it is nearly at the same location on the electronic energy surface as the initial guess, while the DO-GMF calculation climbs up (purple line) to one of the two equivalent second order saddle points (purple diamond), corresponding to a symmetry-broken solution, S$_{2}^{\text{BS}}$. The origin of the contour graph of the electronic energy surface is at the symmetry-pure solution, S$_{2}^{\text{S}}$. The lower graph shows the energy curves for S$_{2}^{\text{BS}}$ (purple curve) and S$_{2}^{\text{S}}$ (red curve). For short H-H  distance, only the symmetry-pure solution exists (orange curve). (b) Sequential calculations of six points spaced by 0.1\,\AA\ along the energy curves shown in (a) using DO-GMF (purple) and DO-MOM (red). The initial guess at the first point at $r_{e}$ is obtained as in (a), while subsequent calculations use the orbitals of the previous H-H distance (indicated by dashed lines). Even after the onset of symmetry breaking at $r_{e} + 0.1$\,\AA, DO-MOM keeps converging on the symmetry-pure solution, S$_{2}^{\text{S}}$, a 1\textsuperscript{st}-order saddle point, while DO-GMF converges consistently on 2\textsuperscript{nd}-order saddle points corresponding to symmetry-broken solutions, S$_{2}^{\text{BS}}$, and thereby provides a more accurate energy curve for atomic dynamics in the excited state.}
    \label{fig:main_h2_optimization_paths}
\end{figure}

The advantage of the DO-GMF method over, for example, DO-MOM is illustrated in Figure \ref{fig:main_h2_optimization_paths}. There, two calculations of the doubly excited state of minimal-basis H\textsubscript{2} using the PBE functional are shown. The first corresponds to the separate point acquisition scheme, and the second corresponds to the sequential one when generating a potential energy surface for atomic dynamics.

In the first calculation, illustrated in Figure \ref{fig:main_h2_optimization_paths}(a), a single point calculation for an H-H distance of 0.4\,Å larger than the minimum energy distance $r_{e}$ is carried out. This distance is beyond the SBO, as can be seen from the contour graph of the electronic energy surface, where the maximum has split up into two maxima. The calculation is initialized using the symmetry-pure ground state solution, S$_{0}^{\text{S}}$, by promoting both electrons from the occupied to the unoccupied orbitals, i.e. 90$^\circ$ orbital rotations, and then a 2\textsuperscript{nd}-order saddle point is targeted. The initial guess turns out to be located at the 1\textsuperscript{st}-order saddle point to within numerical accuracy, thereby corresponding to the symmetric doubly excited state, S$_{2}^{\text{S}}$. Despite the initial close proximity to a 1\textsuperscript{st}-order saddle point, the DO-GMF method climbs up the energy surface, converging on one of the two equivalent 2\textsuperscript{nd}-order saddle points that correspond to symmetry-broken S$_{2}^{\text{BS}}$ solutions and have higher energy than the symmetric S$_{2}^{\text{S}}$ state. However, a DO-MOM calculation using an L-SR1 quasi-Newton optimizer\cite{Levi2020} converges right away on the 1\textsuperscript{st}-order saddle point when starting from this initial guess and is not able to climb up to the 2\textsuperscript{nd}-order saddle point. 

The second example, illustrated in Figure \ref{fig:main_h2_optimization_paths}(b), is a sequential calculation of six points along the potential energy curve for the second excited state of the minimal-basis H\textsubscript{2} molecule. The H-H distance is increased by 0.1\,Å at each step starting from the optimal bond length in the ground electronic state, $r_{e}$. Again, the calculations target a 2\textsuperscript{nd}-order saddle point to find the second excited state. The calculation is initialized by performing 90$^\circ$ orbital rotations from the symmetry-pure ground state solution, S$_{0}^{\text{S}}$ at $r_{e}$. This initial guess is close to the 2\textsuperscript{nd}-order saddle point corresponding to the symmetry-pure solution, S$_{2}^{\text{S}}$, the only doubly excited state solution that exists before the SBO. Both DO-GMF and DO-MOM converge on this stationary point. Then, the H-H distance is incremented to $r_{e} + 0.1$\,\AA\ which is beyond the SBO, and two types of solutions emerge: a 1\textsuperscript{st}-order saddle point corresponding to the symmetry-pure doubly excited state solution, S$_{2}^{\text{S}}$, and a pair of equivalent 2\textsuperscript{nd}-order saddle points corresponding to the spatially symmetry-broken solution, S$_{2}^{\text{BS}}$. The 1\textsuperscript{st}-order saddle point on the energy surface for $r_{e} + 0.1$\,\AA\ is located at the same point on the electronic energy surface as the 2\textsuperscript{nd}-order saddle point on the energy surface for $r_{e}$ before the SBO. Therefore, the initial guess in the calculation for $r_{e} + 0.1$\,\AA\ is located right at the 1\textsuperscript{st}-order saddle on the electronic energy surface, so the DO-MOM calculation converges there. However, DO-GMF is able to move away from the 1\textsuperscript{st}-order saddle point and converges on one of the two 2\textsuperscript{nd}-order saddle points corresponding to a symmetry-broken solution. As the H-H distance is increased further, the two equivalent 2\textsuperscript{nd}-order saddle points move farther away from each other, and DO-GMF moves on the electronic energy surface, while the DO-MOM calculations repeatedly converge on the 1\textsuperscript{st}-order saddle point corresponding to the symmetry-pure solution.

These examples illustrate how convergence to a specific excited state solution can be achieved with DO-GMF, thereby making it possible to generate a potential energy surface for atomic dynamics systematically making use of the advantage symmetry-broken solutions provide. Methods that do not guarantee convergence on a saddle point of a given order have a tendency to converge on the stationary solution closest to the initial guess, thereby producing the less accurate symmetry-pure states.

\section{Results}
\subsection{Energy curve for excited ethylene}
A more challenging example of the way the DO-GMF method can produce a potential energy curve for atomic dynamics in the presence of symmetry breaking is given below in calculations of a doubly excited state of ethylene. At a certain value of the torsional angle, symmetry-broken solutions emerge, and they can be obtained by converging consistently on a 2\textsuperscript{nd}-order saddle point on the electronic energy surface. 

\begin{figure}[!h]
    \centering
    \includegraphics[width=\textwidth]{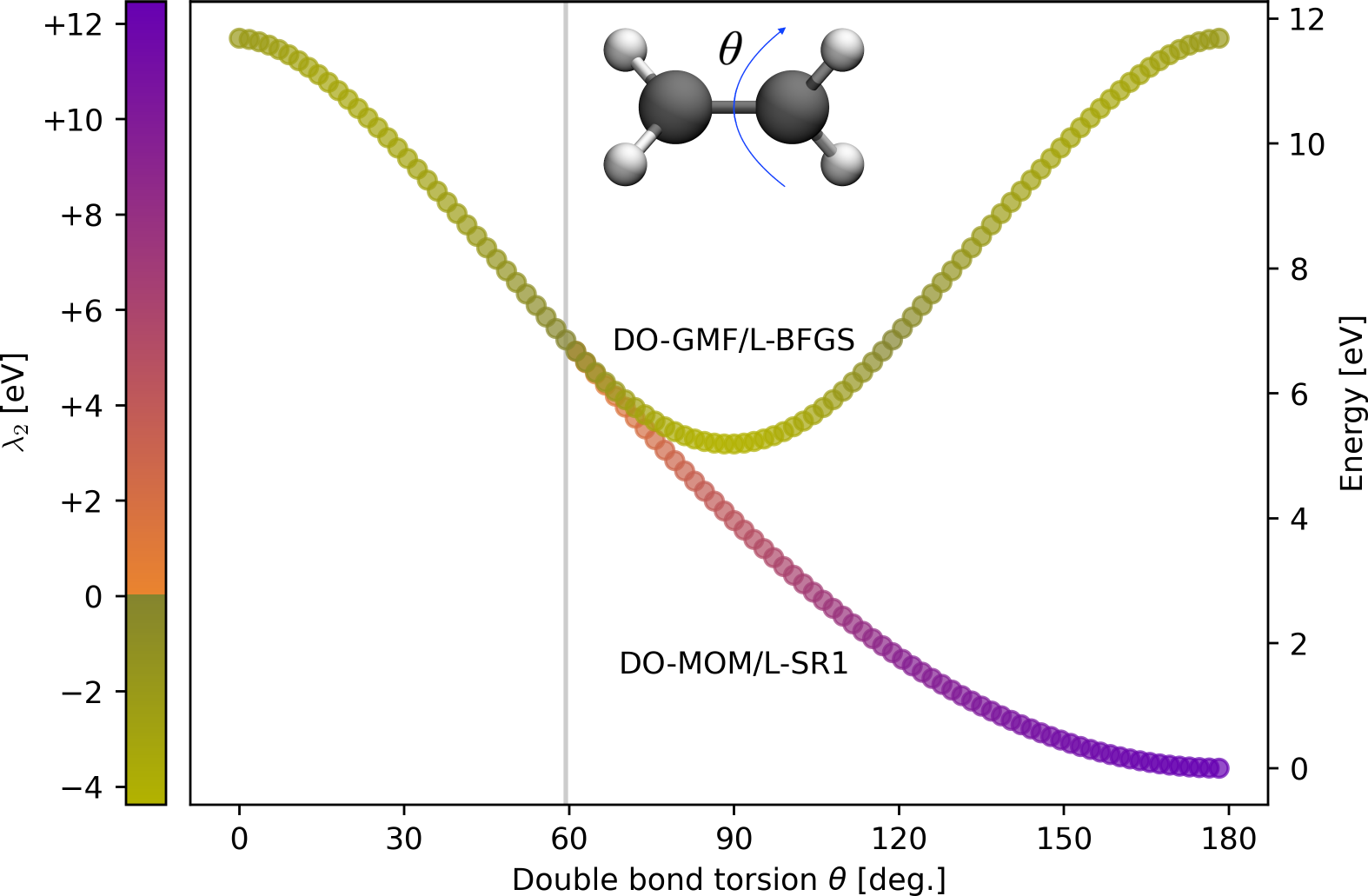}
    \caption{Calculated energy of a doubly excited state of ethylene as a function of the torsional angle $\theta$ calculated with DO-GMF and DO-MOM using sequential point acquisition. A double excitation from the ground state is performed to initialize the excited state calculation at the first geometry, $\theta = 0^\circ$. The DO-GMF calculation targets a 2\textsuperscript{nd}-order saddle point. The points on the curves are colored according to the value of the second eigenvalue of the electronic Hessian, $\lambda_{2}$, while the gray vertical line marks where symmetry-broken solutions appear. Before that, both DO-MOM and DO-GMF converge on the 2\textsuperscript{nd}-order saddle point corresponding to the symmetry-pure solution, $\pi^0\pi^{*2}$. After that, DO-MOM converges on a 1\textsuperscript{st}-order saddle point corresponding to the symmetry-pure solution giving an incorrect potential energy curve. Instead, the DO-GMF calculations keep converging on a 2\textsuperscript{nd}-order saddle point corresponding to a symmetry-broken solution with ionic character (H$_2$C$^+$C$^-$H$_2$/H$_2$C$^-$C$^+$H$_2$), thereby providing a more accurate potential energy curve.}
    \label{fig:main_eigenvalues_ethylene}
\end{figure}

Figure \ref{fig:main_eigenvalues_ethylene} shows the energy curve of ethylene calculated with both the DO-GMF and DO-MOM methods as a function of the C=C torsional angle. 
The potential energy curves are calculated using sequential point acquisition starting from the planar geometry of the ethylene molecule. The excited state calculation at the first geometry is initialized by constructing a double excitation from the ground state solution (90$^\circ$ HOMO-LUMO rotation in both spin channels). The DO-GMF calculations always converge on a 2\textsuperscript{nd}-order saddle point on the electronic energy surface. To characterize the saddle point order of the obtained solutions, Figure \ref{fig:main_eigenvalues_ethylene} shows the value of the second electronic Hessian eigenvalue, $\lambda_{2}$, for each point along the potential energy curve. As the torsional angle is increased towards 90$^\circ$, the double bond in ethylene is broken. At the SBO, $\lambda_{2}$ becomes zero. Thereafter, DO-MOM converges to a symmetry-pure solution with positive $\lambda_{2}$, thereby corresponding to a 1\textsuperscript{st}-order saddle point. The solution obtained with DO-MOM loses another instability at a torsional angle of 120$^{\circ}$, where it coalesces with a ground state solution with broken spin symmetry\cite{Schmerwitz2022}. The symmetry-pure solution obtained using DO-MOM beyond the SBO has covalent character, $\pi^0\pi^{*2}$. This solution gives a diabatic energy surface unlike the energy curve obtained from high-level multireference calculations\cite{Schmerwitz2022, Barbatti2014, Salem1972}, while DO-GMF  converges on a 2\textsuperscript{nd}-order saddle point corresponding to a solution with ionic character (H$_2$C$^+$C$^-$H$_2$/H$_2$C$^-$C$^+$H$_2$), analogous to the S$_2^{\mathrm{BS}}$ solution of minimal-basis H$_2$ shown in Figure \ref{fig:main_h2_pes_plot}. The spatial symmetry is broken, and the potential energy curve displays a minimum at a C=C torsional angle of $\theta = 90^\circ$ in agreement with the multireference results\cite{Salem1972}. Therefore, by converging on a saddle point of the appropriate order using DO-GMF, the calculations can give a qualitatively correct potential energy curve for atomic dynamics.

\subsection{Charge transfer excitations}
Excitations involving large changes in electron density are especially prone to variational collapse and can lead to convergence problems because they often correspond to high-order saddle points on the electronic energy surface, even when the calculation is initialized by an excitation involving ground state orbitals close to the HOMO and LUMO.\cite{Ivanov2021, Levi2020} As a proof of principle, the performance of DO-GMF is first demonstrated in calculations of the open-shell singlet $\pi^{*} \leftarrow \pi'$ charge transfer excited state of nitrobenzene (the notation is taken from ref.\ ~\citenum{Mewes2014}), which has successfully been calculated previously with DO-MOM,\cite{Levi2020} but is known to be problematic for SCF-MOM.\cite{Ivanov2021, Levi2020, Levi2020_2, Hait2020, Mewes2014} Then, a more challenging calculation is presented, a charge transfer excitation in orthogonally twisted N-phenylpyrrole\cite{Loos2021}, where DO-MOM is prone to convergence failure but DO-GMF performs well.

\subsubsection{Nitrobenzene}
\begin{figure}[!h]
    \centering
    \includegraphics[width=\textwidth]{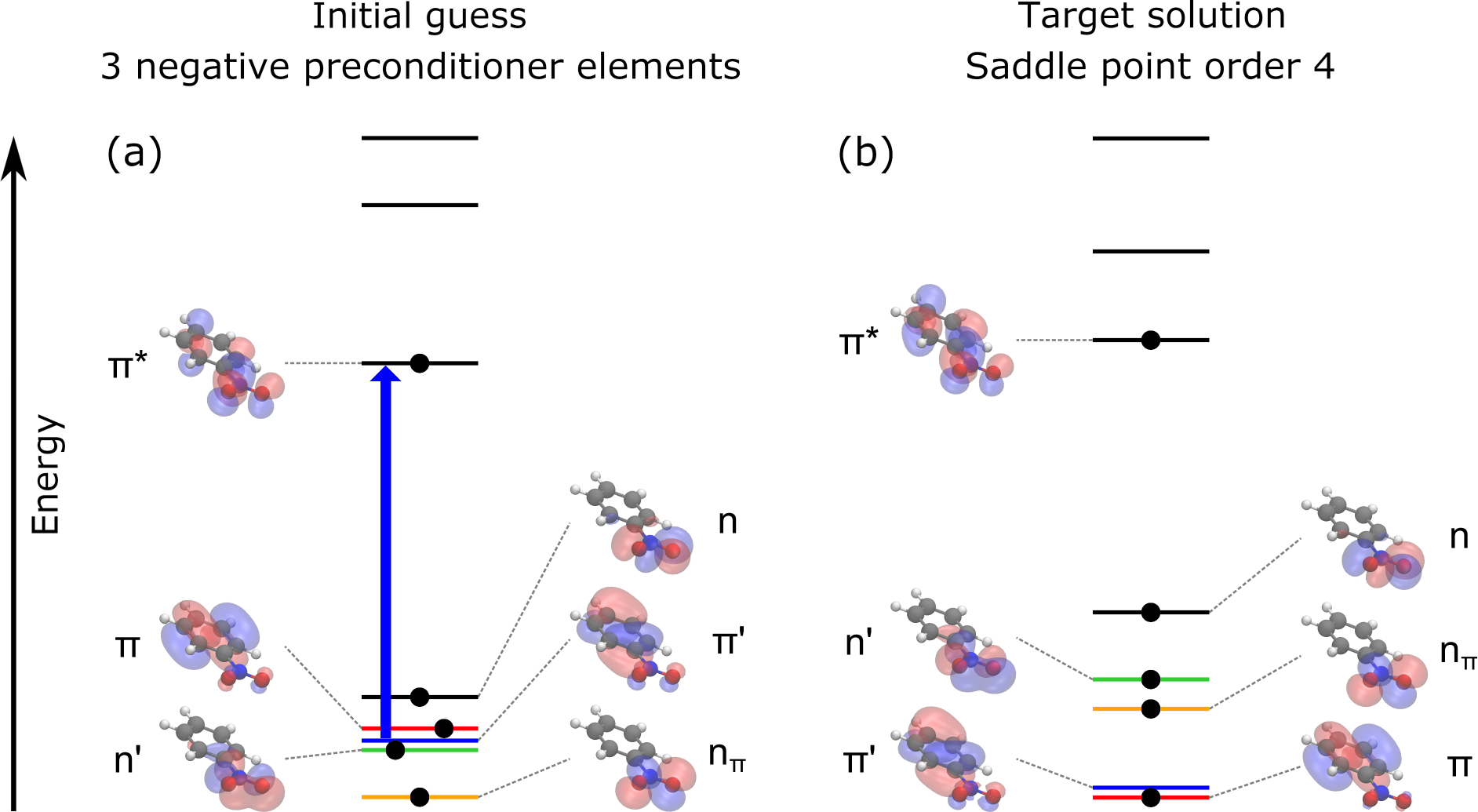}
    \caption{(a) Ground state orbitals of the nitrobenzene molecule with occupation numbers chosen to provide an initial guess for calculations of the $\pi^{*}\leftarrow\pi'$ charge transfer excited state, obtained using the PBE functional and def2-TZVPD+sz basis set. (b) Orbitals of the converged excited state obtained by converging on a 4\textsuperscript{th}-order saddle point on the electronic energy surface. The orbitals are rendered for an isovalue of $\pm0.1\,\mathrm{\AA}^{-1.5}$.}
    \label{fig:main_Nitrobenzene_Orbitals}
\end{figure}

Figure\,\ref{fig:main_Nitrobenzene_Orbitals}(a) shows the orbitals obtained in a ground state calculation of nitrobenzene as well as the change in orbital occupation corresponding to a singlet $\pi^{*} \leftarrow \pi'$ charge transfer excited state. Charge is transferred from the aromatic $\pi$-system to the nitro group.
Based on the ground state orbitals, the excitation is from the HOMO-2 to the LUMO, indicating that an electron hole is formed below three occupied orbitals. The excited state at the initial guess,  before orbital relaxation, thereby, appears to correspond to a 3\textsuperscript{rd}-order saddle point, according to a diagonal approximation of the Hessian. The preconditioner for the quasi-Newton algorithm (eq\,\ref{eq:diagonal_hessian}) indeed has three negative diagonal elements. However, the electronic Hessian at this initial guess based on the ground state orbitals turns out to have 18 negative eigenvalues, many of them being close to zero (five negative eigenvalues are larger than -1\,eV). A better estimate of the saddle point order corresponding to the target excited state is obtained by freezing the HOMO-2 and the LUMO of the ground state and relaxing all other orbitals. The resulting Hessian has five negative eigenvalues, one of them being close to zero (larger than -1\,eV). The charge transfer induces a reordering of the orbitals, as the orbitals that are localized at the phenyl group are stabilized, while orbitals localized at the nitro group are destabilized. The final, converged excited state turns out to correspond to a 4\textsuperscript{th}-order saddle point with the Hessian having four negative eigenvalues. The orbitals for the converged excited state are shown in Figure\,\ref{fig:main_Nitrobenzene_Orbitals}(b). There, the hole is below four occupied orbitals, rather than three as judged from the ground state calculation.

\begin{figure}[!h]
    \centering
    \includegraphics[width=\textwidth]{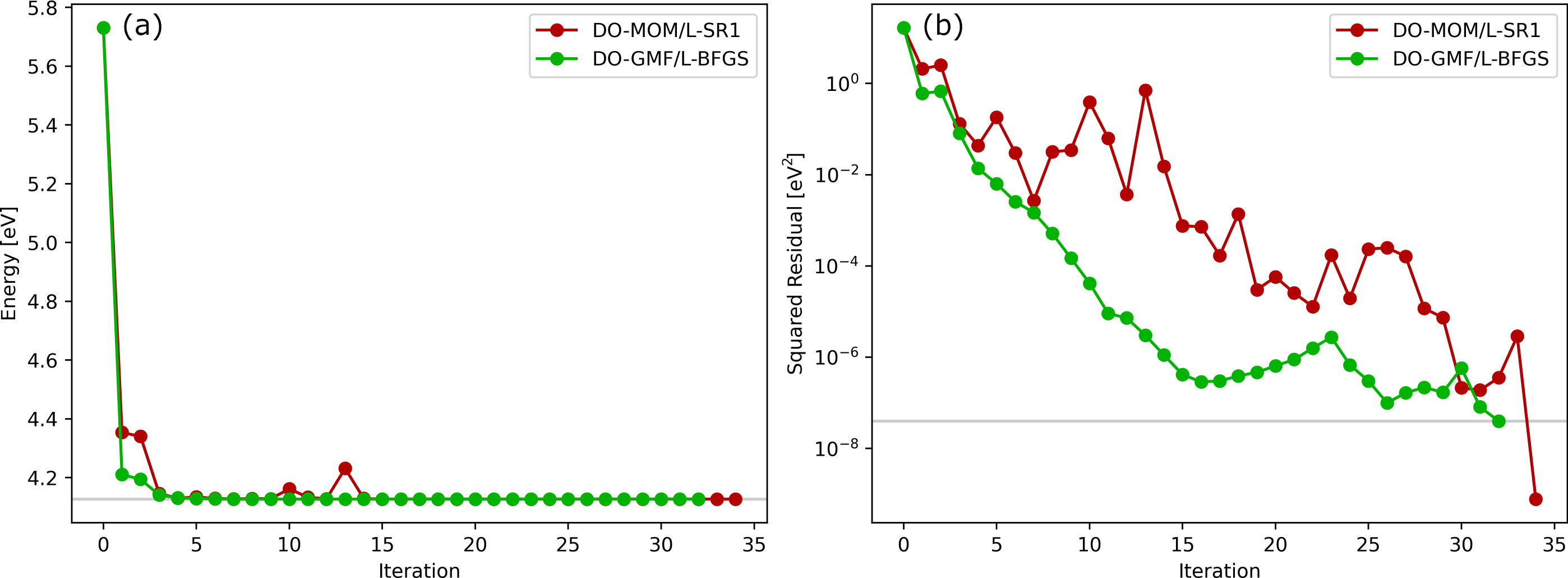}
    \caption{Convergence of DO-MOM and DO-GMF calculations of the open-shell singlet $\pi^{*} \leftarrow \pi'$ charge transfer excited state of nitrobenzene measured in terms of the excitation energy in (a) and squared residual of the KS equations in (b). The calculations use a maximum step length of 0.2 in the quasi-Newton Hessian update (a default setting in GPAW).}
    \label{fig:main_nitrobenzene_convergence}
\end{figure}

Figure \ref{fig:main_nitrobenzene_convergence} shows the convergence of the energy and residual of the KS equations in the DO-MOM and DO-GMF calculations starting from the same initial guess (see also ref \cite{Levi2020} for the DO-MOM calculations). 
DO-MOM with L-SR1 is able to converge on the excited state, meaning that the L-SR1 Hessian update is in this case able to develop an additional negative Hessian eigenvalue over the course of the optimization. DO-GMF using the L-BFGS update and set to target a 4\textsuperscript{th}-order saddle point converges on the same excited state, but converges more smoothly than the DO-MOM calculation and in slightly fewer iterations.

\subsubsection{N-Phenylpyrrole}
\begin{figure}[!h]
    \centering
    \includegraphics[width=\textwidth]{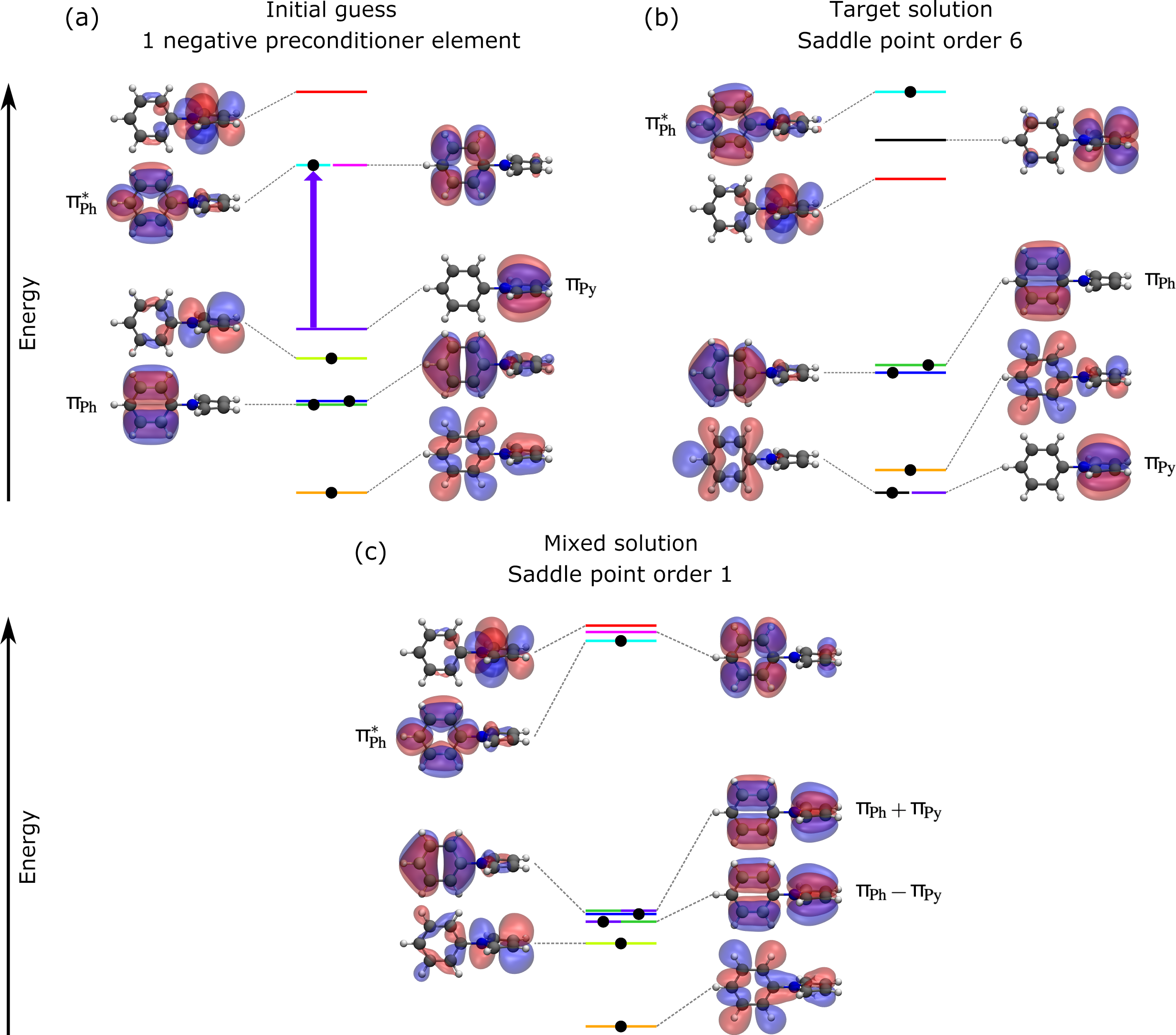}
    \caption{(a) Ground state orbitals of N-phenylpyrrole with changed occupations used as initial guess for DO-MOM and DO-GMF calculations of the $\pi^{*}_{\mathrm{ph}}\leftarrow\pi_{\mathrm{py}}$ charge transfer excited state using the PBE functional and a cc-pVDZ+sz basis set. (b) Orbitals of the target charge transfer solution, which is reached by converging on a 6\textsuperscript{th}-order saddle point. (c) Orbitals of the 1\textsuperscript{st}-order saddle point solution obtained in a calculation with DO-MOM using the default maximum step length of 0.2 for the quasi-Newton Hessian update. All orbitals are shown at an isovalue of $\pm 0.05\,\mathrm{\AA}^{-1.5}$.}
    \label{fig:main_PP_Orbitals}
\end{figure}

A second example of a charge transfer excitation is shown in Figure\,\ref{fig:main_PP_Orbitals}. The promotion of an electron from the HOMO to the LUMO of the ground state wave function of twisted N-phenylpyrrole corresponds to transfer of charge from the pyrrole group ($\pi_{\mathrm{py}}$) to the phenyl group ($\pi^{*}_{\mathrm{ph}}$) as can be seen from the rendering of the orbitals.
Since this excitation appears to be a HOMO to LUMO transition based on the ground state orbitals, the preconditioner at the initial guess has only one negative element. Yet, the electronic Hessian has 28 negative eigenvalues, with five being close to zero (larger than -1\,eV). A relaxation of the orbitals with the HOMO and LUMO frozen leads to a Hessian with seven negative eigenvalues, one being close to zero (larger than -1\,eV), indicating that the saddle point on the energy surface corresponding to the converged excited state is of  6\textsuperscript{th} order. This prediction turns out to be correct, as the hole gets stabilized during the variational optimization of the orbitals dropping below three other occupied orbitals in energy, and furthermore, two other unoccupied orbitals get stabilized with respect to the orbital to which the electron is excited. The converged excited state has 6 negative Hessian eigenvalues, thereby corresponding to a 6\textsuperscript{th}-order saddle point. 

\begin{figure}[!h]
    \centering
    \includegraphics[width=\textwidth]{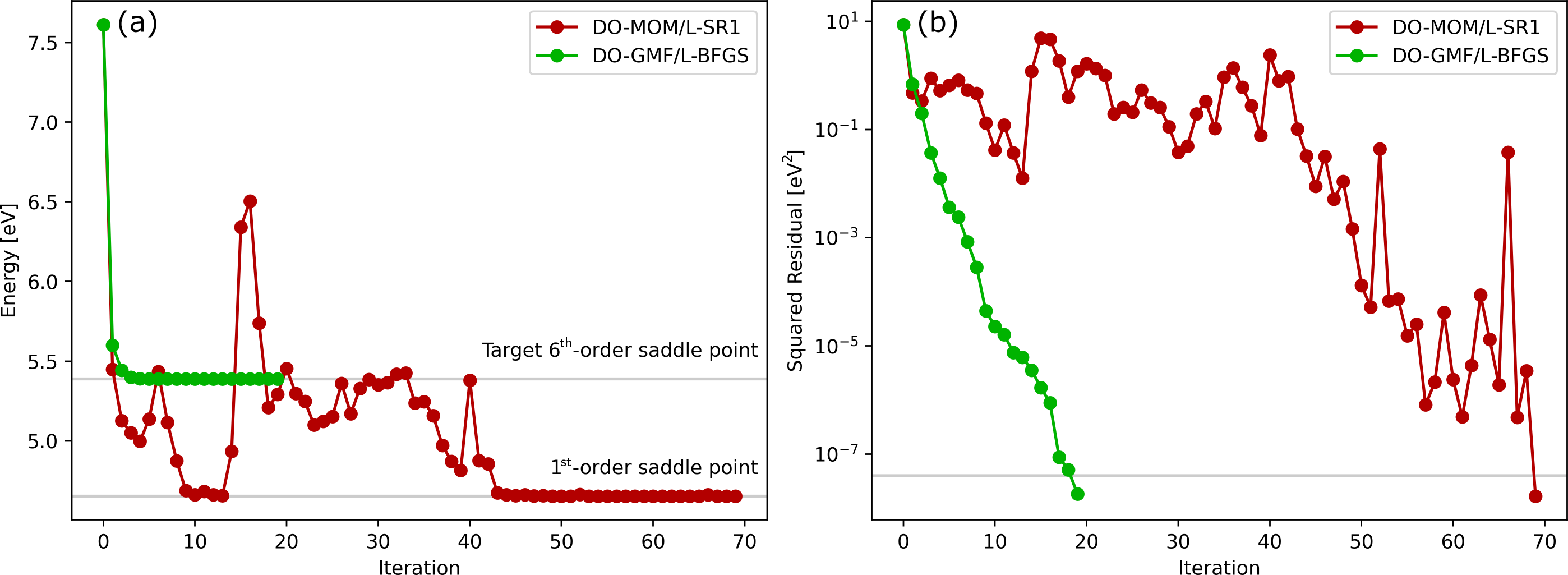}
    \caption{Convergence of the excitation energy (a) and squared residual of the KS equations (b) during DO-MOM and DO-GMF calculations of the open-shell singlet $\pi^{*}_{\mathrm{ph}}\leftarrow\pi_{\mathrm{py}}$ charge transfer excited state of N-phenylpyrrole. Both DO-MOM and DO-GMF calculations use a maximum step length of 0.2 for the quasi-Newton Hessian update (default in GPAW). While DO-MOM displays erratic convergence behavior and ultimately collapses to a lower-energy 1\textsuperscript{st}-order saddle point solution with small dipole moment, DO-GMF rapidly converges to the target 6\textsuperscript{th}-order saddle point solution.}
    \label{fig:main_PP_convergence}
\end{figure}

Figure \ref{fig:main_PP_convergence} shows the rate of convergence of the DO-GMF calculation and a comparison with a DO-MOM calculation starting from the same initial guess for the $\pi^{*}_{\mathrm{ph}}\leftarrow\pi_{\mathrm{py}}$ charge transfer excitation.
DO-GMF converges in a robust way in only 19 optimization steps and with a monotonically decreasing energy and squared residual of the KS equations. The DO-MOM calculation, however, using the default maximum step length of 0.2 collapses to a 1\textsuperscript{st}-order saddle point and fails to converge on the targeted charge transfer solution. This collapse is a consequence of the mismatch between the number of concave directions estimated by the preconditioner and that found at the converged charge transfer solution and an inability of the L-SR1 update to develop the missing negative Hessian eigenvalues. The orbitals of the 1\textsuperscript{st}-order saddle point solution obtained with DO-MOM are visualized in Figure \ref{fig:main_PP_Orbitals}c. The frontier orbitals contain a pair of occupied-unoccupied orbitals delocalized over the entire molecule, which leads to a small dipole moment. These delocalized orbitals arise from $\sim$45$^{\circ}$ mixing between the $\pi_{\mathrm{py}}$ hole localized on the pyrrole group and a $\pi_{\mathrm{ph}}$ occupied orbital localized on the phenyl group, giving linear combination orbitals with characters $\frac{1}{\sqrt{2}}(\pi_{\mathrm{py}}+\pi_{\mathrm{ph}})$ and $\frac{1}{\sqrt{2}}(\pi_{\mathrm{py}}-\pi_{\mathrm{ph}})$\,. MOM by construction cannot prevent variational collapse for rotational angles smaller than 45$^{\circ}$. 

\subsection{Bond dissociation in AgPtPOP}
\begin{figure}[!h]
    \centering
    \includegraphics{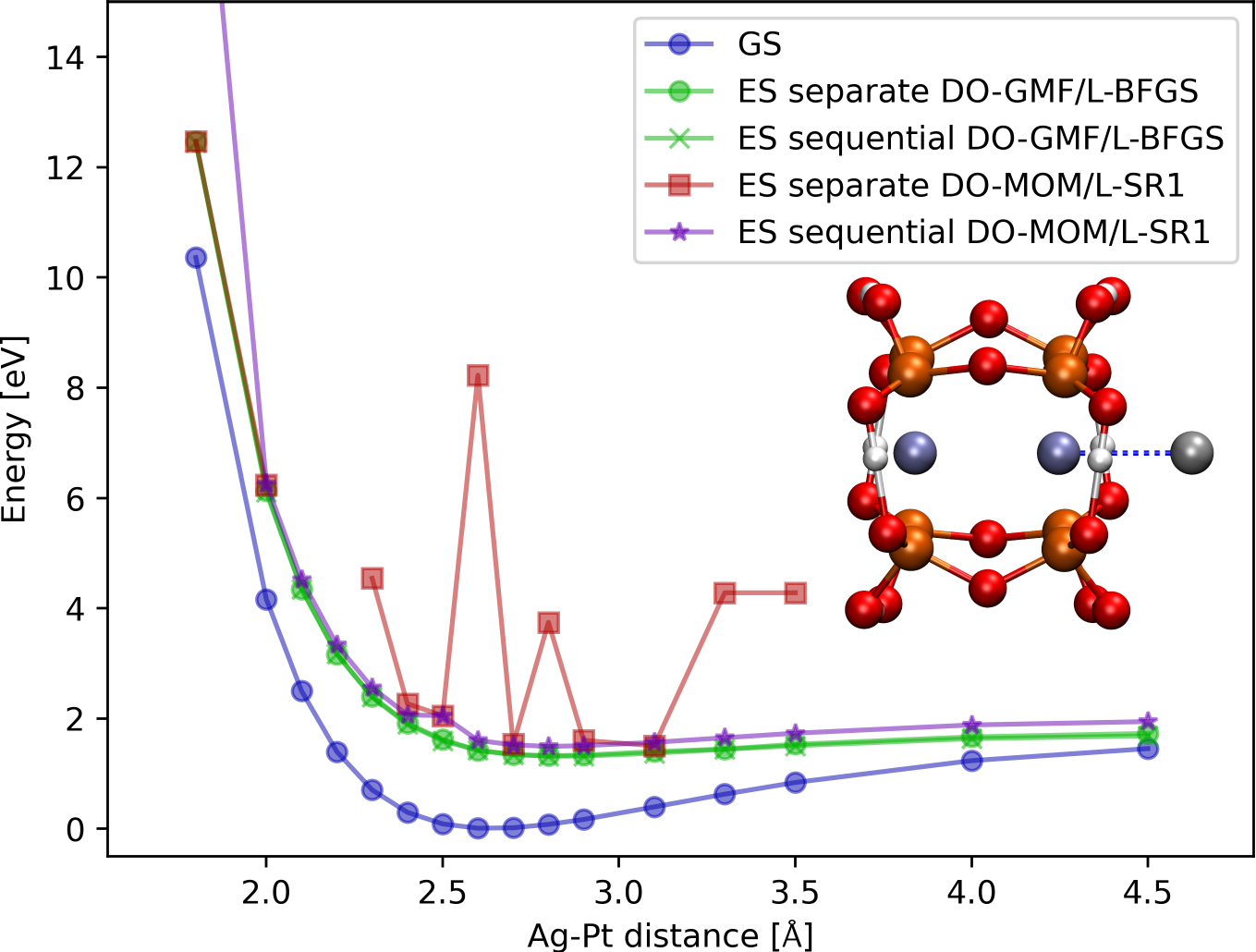}
    \caption{Energy of the AgPtPOP complex in the ground state (GS, blue) and in the singlet excited states (ES, green) corresponding to a LUMO$\leftarrow$HOMO excitation as a function of the Ag-Pt distance. The excited state points are obtained using DO-GMF by targeting a 1\textsuperscript{st}-order saddle point, using either a sequential (green crosses) or separate acquisition (green circles). Results of DO-MOM calculations (purple stars for the sequential acquisition and red squares for the separate acquisition) are also shown. When the points are calculated separately, the initial guess for an excited state calculation at a given point is given by the orbitals of the ground state solution obtained for the same Ag-Pt distance. In the sequential calculations, the initial guess at a given point is given by the excited state orbitals obtained at the previous point, starting from an Ag-Pt distance of 2.5\,\AA\ and then increasing and decreasing the distance. The complex is visualized in the inset, showing the relevant Ag-Pt distance with a dashed line, and the color code for the atoms being: Ag = silver, Pt = blue, P = orange, O = red, H = white. The calculations use the BLYP functional and a dzp basis set of numeric atomic orbitals.}
    \label{fig:main_AgPtPOP}
\end{figure}

As an example of an application to a larger molecule, Figure\,\ref{fig:main_AgPtPOP} shows the results of a calculation for a diplatinum and silver complex, AgPtPOP.
This compound derives from binding of an Ag$^+$ ion with the d$^8$-d$^8$ diplatinum complex [Pt\textsubscript{2}(P\textsubscript{2}O\textsubscript{5}H\textsubscript{2})\textsubscript{4}]\textsuperscript{4–} (PtPOP)\cite{Gray2017}. The latter has attracted considerable interest in recent years due to its photocatalytic properties as well as rich photophysics, which has led to fundamental studies of photoinduced bond vibration dynamics and energy relaxation\cite{Haldrup2019, Monni2018, Levi2018, Veen2011}. Upon photoexcitation, an electron is promoted from an antibonding d$\sigma^*$ HOMO to a bonding p$\sigma$ LUMO metal-metal orbital, leading to Pt-Pt contraction and subsequent coherent vibrations. Formation of an exciplex with an additional metal ion, as in AgPtPOP\cite{Clodfelter1994, Christensen2010}, gives an opportunity to obtain insights into metal-metal interactions and dynamics by excitation to the lowest excited state. Computational studies of these dynamical processes require the potential energy surface corresponding to the lowest excited electronic state.

Figure\,\ref{fig:main_AgPtPOP} shows the calculated energy for the ground state and open-shell singlet excited state corresponding to HOMO-LUMO excitation as a function of the Ag-Pt distance. The Ag atom is placed along the Pt-Pt axis, and the potential energy curves are generated using either a separate point acquisition scheme where an excited state calculation for a given Ag-Pt distance is initialized using the ground state orbitals at that distance, or by a sequential scheme where the excited state calculation for an Ag-Pt distance of 2.5\,\AA\ is initialized using the ground state orbitals, and then, the converged excited state orbitals of the previous distance are used as the initial guess for the next calculation, both increasing the Ag-Pt distance and decreasing it. The DO-GMF calculations of the excited state are set to target a 1\textsuperscript{st}-order saddle point and they produce a smooth curve, with either acquisition scheme, converging to the same solution for a given point on the curve. However, the DO-MOM calculations with the separate acquisition scheme converge on several different solutions as evidenced by the saddle point orders ranging from 1 to larger than 25, resulting in a discontinuous energy curve. Also, a few points on the curve could not be converged within 500 optimization steps. With the sequential point acquisition scheme, DO-MOM converges to a consistent solution where the saddle point order is 3 with the exception of the first point on the curve which corresponds to a 2\textsuperscript{nd}-order saddle point. 

\section{Discussion}
The DO-GMF method presented here is found to be reliable and robust for variational density functional calculations of excited electronic states by converging on a saddle point of a specified order on the electronic energy surface. The method has the distinct advantage that no precaution is needed to prevent collapse to the ground state, such as MOM. It outperforms the DO-MOM method which has, in turn, previously been shown to have superior convergence properties than commonly used SCF-MOM approaches\cite{Schmerwitz2022, Ivanov2021, Levi2020}. 

Variational excited state calculations have a tendency to converge on the stationary solution closest to the initial guess, irrespective of what saddle point order it corresponds to. Therefore, if multiple  electronic states emerge as an atomic configuration is changed slightly, possibly breaking symmetries, the variational calculation tends to converge on the state that is closest to the initial guess, thereby conserving symmetry. This behavior can lead to unphysical potential energy curves for atomic dynamics and give incorrect excited states, as exemplified by the calculations of bond breaking in the H$_2$ and ethylene molecules. DO-GMF is able to follow solutions that move away from the original position on the electronic energy surface as the atomic configuration changes, as the saddle point order of the converged solution is conserved. This property is illustrated by the calculations of the symmetry-broken doubly excited states of the H$_2$ and ethylene molecules. The potential energy curves calculated with DO-GMF are in qualitative agreement with multireference results.

The DO-GMF calculation, however, requires as input the saddle point order of the targeted excited states. The diagonal approximation to the Hessian using the ground state orbitals with non-aufbau occupation numbers tends to underestimate the saddle point order for excitations with large rearrangement of the orbitals, as exemplified by the charge transfer excited states of nitrobenzene and N-phenylpyrrole. Also, eigendecomposition of the full electronic Hessian at the initial guess based on the ground state orbitals does not provide the correct number of negative eigenvalues, as the electronic energy surface at the initial guess does not have the same curvature as at the converged excited state. In the cases of nitrobenzene and N-phenylpyrrole, the number of negative eigenvalues obtained this way is much larger than the saddle point order of the converged excited state. A constrained optimization where the orbitals corresponding to the electron-hole pairs are fixed is found to provide a better estimate of the saddle point order due to the partial inclusion of relaxation effects. In the cases of nitrobenzene and N-phenylpyrrole, the saddle point order estimated this way overestimates the true saddle point order by only one, due to a negative eigenvalue that is close to zero (larger than -1 eV). An improved strategy could be to re-evaluate the diagonal approximation of the Hessian after the first few steps of the wave function optimization, when the orbitals have started rearranging in response to the new electron distribution. A localization transformation of the initial orbitals may also be useful. Future efforts should aim to assess and compare these possible strategies to estimate the saddle point order of the target excited state.

The DO-MOM method would also benefit from a preconditioner that accurately estimates the degrees of freedom along which the energy needs to be maximized. If the relaxed solution is a saddle point of higher order than estimated by the diagonal Hessian approximation at the initial guess, the DO-MOM optimization can start collapsing to a lower-order saddle point before the quasi-Newton algorithm is able to develop additional negative eigenvalues, as seen in the calculations of the charge transfer excitation in N-phenylpyrrole. There, MOM cannot avoid the variational collapse as the orbitals in the lower-energy solution mix by less than 45$^\circ$. This failure shows that the commonly used MOM method does not guarantee that variational collapse is prevented. However, DO-GMF is less dependent on the preconditioner for the quasi-Newton step, and variational collapse is excluded by construction, thus convergence on a saddle point of the target order is guaranteed. 

The generalized Davidson method used in DO-GMF requires an initial guess for each target eigenvector. At the first wave function optimization step, the most straightforward guess consists of the frontier orbital rotations up to the target saddle point order in order of increasing corresponding element of the diagonal Hessian approximation. However, the electronic Hessian eigenvectors often contain significant contributions from multiple orbital rotations. Therefore, such an initial guess may be far from the actual Hessian eigenvectors. Subsequent partial Hessian diagonalizations typically converge faster since the converged eigenvectors of the previous optimization step can be used as the initial guess. In order to accelerate the first partial Hessian diagonalization, it could be worthwhile to use a generalized version of the dimer method to converge more than one of the lowest eigenpairs at the same time. Such a generalized dimer method would optimize multiple orthogonal dimers at the same time, meaning that multiple optimal dimer rotations are found simultaneously. While the lack of a preconditioner for a generalized dimer method could make it difficult to ensure convergence to the lowest eigenvectors, the possibility of using the L-BFGS optimizer to find the optimal dimer rotations could accelerate convergence significantly for the first partial diagonalization step.

The computational effort of DO-GMF is greater than that of DO-MOM as DO-MOM requires one energy/gradient evaluation per wave function optimization step, while DO-GMF requires one additional energy/gradient calculation per target eigenpair per step if a forward finite difference approximation is used in the partial Hessian diagonalization. As the accuracy of the forward finite difference approximation scales linearly in the finite difference step size, the latter should be chosen as small as possible, while ensuring numerical stability. A finite difference step size of $10^{-3}$ has been found to be numerically stable and accurate enough for the calculations presented here. Quadratic scaling can be obtained by using a central finite difference approximation performing twice the amount of energy/gradient evaluations, but we have not found it necessary in the calculations presented here.

\section{Conclusion}
The DO-GMF method is presented for variational calculations of excited electronic states where a stationary solution is reached by converging on a saddle point of a given order. The method involves a generalization of the minimum mode following approach where an $n$\textsuperscript{th}-order saddle point is found by inverting the components of the gradient in the direction of the eigenvectors of the $n$ lowest eigenvalues of the Hessian. The method recasts the saddle point search as a minimization using the modified gradient, so variational collapse to the ground state is prevented by construction. An implementation is presented where the exponential transformation direct optimization is used in combination with the L-BFGS algorithm to perform the orbital optimization and the generalized Davidson method is used for the partial diagonalization of the electronic Hessian. The performance of DO-GMF is demonstrated in calculations using generalized gradient approximation density functionals of excited states that are challenging for a previously proposed method based on direct optimization and the maximum overlap method. DO-GMF is demonstrated to track excited states through atomic configurations where the symmetry of the wave function breaks in calculations of the potential energy curves of H\textsubscript{2} and ethylene by systematically targeting a saddle point of a given order. The method has, furthermore, been shown to converge in a robust way in calculations of challenging charge transfer excited states of nitrobenzene and N-phenylpyrrole as well as the lowest excited state of a large diplatinum and silver complex, AgPtPOP, as the Ag-Pt distance is varied. 

The results presented here indicate that DO-GMF is a promising tool for Born-Oppenheimer and non-adiabatic simulations of atomic dynamics involving sequential excited state calculations. Additional tests on more complicated systems are, of course, needed to verify that convergence on a saddle point of a given order can be used as a guiding principle when symmetry-broken solutions provide improved estimate of the energy of the system. Another issue is that symmetry-broken solutions introduce undesirable features, such as spurious dipole and magnetic moments, which can, in turn, lead to inaccurate atomic forces along degrees of freedom other than bond breaking coordinates. Quite likely, their symmetry-related counterparts will need to be taken into account to accurately evaluate atomic forces over the entire energy surface.

\begin{acknowledgement}
This work was supported by the Icelandic Research Fund (grant agreements nos. 217751, 217734, 196070). The calculations were carried out at the Icelandic High Performance Computing Center (IHPC). The authors thank Asmus O. Dohn, Christoffer H. Egeberg and Kristoffer Haldrup for useful discussions and support with the calculations on the AgPtPOP complex, and Elli Selenius for support with the calculations on twisted N-phenylpyrrole.
\end{acknowledgement}

\begin{suppinfo}
The authors confirm that the data supporting the findings of this study are available within the article and/or its supplementary materials.

Second Hessian eigenvalue along doubly excited state energy curve of H\textsubscript{2}; Convergence of the $\pi_{\mathrm{ph}}^{*} \leftarrow \pi_{\mathrm{py}}$ charge transfer excited state of twisted N-phenylpyrrole with DO-MOM and DO-GMF when a smaller optimization step size of 0.1 is used; Optimized geometries of all systems.
\end{suppinfo}

\bibliography{main_References}

\providecommand{\latin}[1]{#1}
\makeatletter
\providecommand{\doi}
  {\begingroup\let\do\@makeother\dospecials
  \catcode`\{=1 \catcode`\}=2 \doi@aux}
\providecommand{\doi@aux}[1]{\endgroup\texttt{#1}}
\makeatother
\providecommand*\mcitethebibliography{\thebibliography}
\csname @ifundefined\endcsname{endmcitethebibliography}
  {\let\endmcitethebibliography\endthebibliography}{}
\begin{mcitethebibliography}{117}
\providecommand*\natexlab[1]{#1}
\providecommand*\mciteSetBstSublistMode[1]{}
\providecommand*\mciteSetBstMaxWidthForm[2]{}
\providecommand*\mciteBstWouldAddEndPuncttrue
  {\def\EndOfBibitem{\unskip.}}
\providecommand*\mciteBstWouldAddEndPunctfalse
  {\let\EndOfBibitem\relax}
\providecommand*\mciteSetBstMidEndSepPunct[3]{}
\providecommand*\mciteSetBstSublistLabelBeginEnd[3]{}
\providecommand*\EndOfBibitem{}
\mciteSetBstSublistMode{f}
\mciteSetBstMaxWidthForm{subitem}{(\alph{mcitesubitemcount})}
\mciteSetBstSublistLabelBeginEnd
  {\mcitemaxwidthsubitemform\space}
  {\relax}
  {\relax}

\bibitem[Runge and Gross(1984)Runge, and Gross]{Runge1984}
Runge,~E.; Gross,~E. K.~U. Density-Functional Theory for Time-Dependent
  Systems. \emph{Phys. Rev. Lett.} \textbf{1984}, \emph{52}, 997\relax
\mciteBstWouldAddEndPuncttrue
\mciteSetBstMidEndSepPunct{\mcitedefaultmidpunct}
{\mcitedefaultendpunct}{\mcitedefaultseppunct}\relax
\EndOfBibitem
\bibitem[Casida(1995)]{Casida1995}
Casida,~M.~E. \emph{Recent Advances in Computational Chemistry}; 1995; pp
  155--192\relax
\mciteBstWouldAddEndPuncttrue
\mciteSetBstMidEndSepPunct{\mcitedefaultmidpunct}
{\mcitedefaultendpunct}{\mcitedefaultseppunct}\relax
\EndOfBibitem
\bibitem[Herbert(2022)]{Herbert2022}
Herbert,~J.~M. Density Functional Theory for Electronic Excited States. 2022;
  \url{https://arxiv.org/abs/2204.10135}\relax
\mciteBstWouldAddEndPuncttrue
\mciteSetBstMidEndSepPunct{\mcitedefaultmidpunct}
{\mcitedefaultendpunct}{\mcitedefaultseppunct}\relax
\EndOfBibitem
\bibitem[Dreuw and Head-Gordon(2004)Dreuw, and Head-Gordon]{Dreuw2004}
Dreuw,~A.; Head-Gordon,~M. Failure of Time-Dependent Density Functional Theory
  for Long-Range Charge-Transfer Excited States: The
  Zincbacteriochlorin-Bacteriochlorin and Bacteriochlorophyll-Spheroidene
  Complexes. \emph{J. Am. Chem. Soc.} \textbf{2004}, \emph{126},
  4007--4016\relax
\mciteBstWouldAddEndPuncttrue
\mciteSetBstMidEndSepPunct{\mcitedefaultmidpunct}
{\mcitedefaultendpunct}{\mcitedefaultseppunct}\relax
\EndOfBibitem
\bibitem[Dreuw and Head-Gordon(2005)Dreuw, and Head-Gordon]{Dreuw2005}
Dreuw,~A.; Head-Gordon,~M. Single-reference ab initio methods for the
  calculation of excited states of large molecules. \emph{Chem. Rev.}
  \textbf{2005}, \emph{105}, 4009--4037\relax
\mciteBstWouldAddEndPuncttrue
\mciteSetBstMidEndSepPunct{\mcitedefaultmidpunct}
{\mcitedefaultendpunct}{\mcitedefaultseppunct}\relax
\EndOfBibitem
\bibitem[Hait and Head-Gordon(2021)Hait, and Head-Gordon]{Hait2021}
Hait,~D.; Head-Gordon,~M. Orbital Optimized Density Functional Theory for
  Electronic Excited States. \emph{J. Phys. Chem. Lett.} \textbf{2021},
  \emph{12}, 4517--4529\relax
\mciteBstWouldAddEndPuncttrue
\mciteSetBstMidEndSepPunct{\mcitedefaultmidpunct}
{\mcitedefaultendpunct}{\mcitedefaultseppunct}\relax
\EndOfBibitem
\bibitem[Hait \latin{et~al.}(2019)Hait, Rettig, and Head-Gordon]{Hait2019}
Hait,~D.; Rettig,~A.; Head-Gordon,~M. Well-behaved versus ill-behaved density
  functionals for single bond dissociation: Separating success from disaster
  functional by functional for stretched H\textsubscript{2}. \emph{J. Chem.
  Phys.} \textbf{2019}, \emph{150}\relax
\mciteBstWouldAddEndPuncttrue
\mciteSetBstMidEndSepPunct{\mcitedefaultmidpunct}
{\mcitedefaultendpunct}{\mcitedefaultseppunct}\relax
\EndOfBibitem
\bibitem[Hait \latin{et~al.}(2019)Hait, Rettig, and Head-Gordon]{Hait2019_2}
Hait,~D.; Rettig,~A.; Head-Gordon,~M. Beyond the Coulson-Fischer point:
  Characterizing single excitation CI and TDDFT for excited states in single
  bond dissociations. \emph{Phys. Chem. Chem. Phys.} \textbf{2019}, \emph{21},
  21761--21775\relax
\mciteBstWouldAddEndPuncttrue
\mciteSetBstMidEndSepPunct{\mcitedefaultmidpunct}
{\mcitedefaultendpunct}{\mcitedefaultseppunct}\relax
\EndOfBibitem
\bibitem[Barca \latin{et~al.}(2018)Barca, Gilbert, and Gill]{Barca2018}
Barca,~G.~M.; Gilbert,~A.~T.; Gill,~P.~M. Simple Models for Difficult
  Electronic Excitations. \emph{J. Chem. Theory Comput.} \textbf{2018},
  \emph{14}, 1501--1509\relax
\mciteBstWouldAddEndPuncttrue
\mciteSetBstMidEndSepPunct{\mcitedefaultmidpunct}
{\mcitedefaultendpunct}{\mcitedefaultseppunct}\relax
\EndOfBibitem
\bibitem[Levine \latin{et~al.}(2006)Levine, Ko, Quenneville, and
  Martínez]{Levine2006}
Levine,~B.~G.; Ko,~C.; Quenneville,~J.; Martínez,~T.~J. Conical intersections
  and double excitations in time-dependent density functional theory.
  \emph{Mol. Phys.} \textbf{2006}, \emph{104}, 1039--1051\relax
\mciteBstWouldAddEndPuncttrue
\mciteSetBstMidEndSepPunct{\mcitedefaultmidpunct}
{\mcitedefaultendpunct}{\mcitedefaultseppunct}\relax
\EndOfBibitem
\bibitem[Athavale \latin{et~al.}(2021)Athavale, Teh, and
  Subotnik]{Athavale2021}
Athavale,~V.; Teh,~H.-H.; Subotnik,~J. On The Inclusion of One Double Within
  CIS and TD-DFT. \emph{J. Chem. Phys.} \textbf{2021}, \emph{155}, 154105\relax
\mciteBstWouldAddEndPuncttrue
\mciteSetBstMidEndSepPunct{\mcitedefaultmidpunct}
{\mcitedefaultendpunct}{\mcitedefaultseppunct}\relax
\EndOfBibitem
\bibitem[Teh and Subotnik(2019)Teh, and Subotnik]{Teh2019}
Teh,~H.~H.; Subotnik,~J.~E. The Simplest Possible Approach for Simulating
  S\textsubscript{0}-S\textsubscript{1} Conical Intersections with DFT/TDDFT:
  Adding One Doubly Excited Configuration. \emph{J. Phys. Chem. Lett.}
  \textbf{2019}, \emph{10}, 3426--3432\relax
\mciteBstWouldAddEndPuncttrue
\mciteSetBstMidEndSepPunct{\mcitedefaultmidpunct}
{\mcitedefaultendpunct}{\mcitedefaultseppunct}\relax
\EndOfBibitem
\bibitem[Shu \latin{et~al.}(2017)Shu, Parker, and Truhlar]{shu2017dual}
Shu,~Y.; Parker,~K.~A.; Truhlar,~D.~G. Dual-Functional Tamm-Dancoff
  Approximation with Self-Interaction-Free Orbitals: Vertical Excitation
  Energies and Potential Energy Surfaces near an Intersection Seam. \emph{J.
  Phys. Chem.} \textbf{2017}, \emph{121}, 9728--9735\relax
\mciteBstWouldAddEndPuncttrue
\mciteSetBstMidEndSepPunct{\mcitedefaultmidpunct}
{\mcitedefaultendpunct}{\mcitedefaultseppunct}\relax
\EndOfBibitem
\bibitem[Shu \latin{et~al.}(2017)Shu, Parker, and Truhlar]{Shu2017}
Shu,~Y.; Parker,~K.~A.; Truhlar,~D.~G. Dual-Functional Tamm–Dancoff
  Approximation: A Convenient Density Functional Method that Correctly
  Describes S\textsubscript{1}/S\textsubscript{0} Conical Intersections.
  \emph{J. Phys. Chem. Lett.} \textbf{2017}, \emph{8}, 2107--2112\relax
\mciteBstWouldAddEndPuncttrue
\mciteSetBstMidEndSepPunct{\mcitedefaultmidpunct}
{\mcitedefaultendpunct}{\mcitedefaultseppunct}\relax
\EndOfBibitem
\bibitem[Shao \latin{et~al.}(2003)Shao, Head-Gordon, and Krylov]{Shao2003}
Shao,~Y.; Head-Gordon,~M.; Krylov,~A.~I. The spin-flip approach within
  time-dependent density functional theory: Theory and applications to
  diradicals. \emph{J. Chem. Phys.} \textbf{2003}, \emph{118}, 4807--4818\relax
\mciteBstWouldAddEndPuncttrue
\mciteSetBstMidEndSepPunct{\mcitedefaultmidpunct}
{\mcitedefaultendpunct}{\mcitedefaultseppunct}\relax
\EndOfBibitem
\bibitem[Yang \latin{et~al.}(2016)Yang, Shen, Zhang, and Yang]{Yang2016}
Yang,~Y.; Shen,~L.; Zhang,~D.; Yang,~W. Conical Intersections from
  Particle-Particle Random Phase and Tamm-Dancoff Approximations. \emph{J.
  Phys. Chem. Lett.} \textbf{2016}, \emph{7}, 2407--2411\relax
\mciteBstWouldAddEndPuncttrue
\mciteSetBstMidEndSepPunct{\mcitedefaultmidpunct}
{\mcitedefaultendpunct}{\mcitedefaultseppunct}\relax
\EndOfBibitem
\bibitem[Bannwarth \latin{et~al.}(2020)Bannwarth, Yu, Hohenstein, and
  Martínez]{Bannwarth2020}
Bannwarth,~C.; Yu,~J.~K.; Hohenstein,~E.~G.; Martínez,~T.~J. Hole-hole
  Tamm-Dancoff-approximated density functional theory: A highly efficient
  electronic structure method incorporating dynamic and static correlation.
  \emph{J. Chem. Phys.} \textbf{2020}, \emph{153}, 024110\relax
\mciteBstWouldAddEndPuncttrue
\mciteSetBstMidEndSepPunct{\mcitedefaultmidpunct}
{\mcitedefaultendpunct}{\mcitedefaultseppunct}\relax
\EndOfBibitem
\bibitem[Ottochian \latin{et~al.}(2020)Ottochian, Morgillo, Ciofini, Frisch,
  Scalmani, and Adamo]{Ottochian2020}
Ottochian,~A.; Morgillo,~C.; Ciofini,~I.; Frisch,~M.~J.; Scalmani,~G.;
  Adamo,~C. Double hybrids and time-dependent density functional theory: An
  implementation and benchmark on charge transfer excited states. \emph{J.
  Comput. Chem.} \textbf{2020}, \emph{41}, 1242--1251\relax
\mciteBstWouldAddEndPuncttrue
\mciteSetBstMidEndSepPunct{\mcitedefaultmidpunct}
{\mcitedefaultendpunct}{\mcitedefaultseppunct}\relax
\EndOfBibitem
\bibitem[Brémond \latin{et~al.}(2021)Brémond, Ottochian, Pérez-Jiménez,
  Ciofini, Scalmani, Frisch, Sancho-García, and Adamo]{Bremond2021}
Brémond,~E.; Ottochian,~A.; Pérez-Jiménez,~A.~J.; Ciofini,~I.; Scalmani,~G.;
  Frisch,~M.~J.; Sancho-García,~J.~C.; Adamo,~C. Assessing challenging intra-
  and inter-molecular charge-transfer excitations energies with double-hybrid
  density functionals. \emph{J. Comput. Chem.} \textbf{2021}, \emph{42},
  970--981\relax
\mciteBstWouldAddEndPuncttrue
\mciteSetBstMidEndSepPunct{\mcitedefaultmidpunct}
{\mcitedefaultendpunct}{\mcitedefaultseppunct}\relax
\EndOfBibitem
\bibitem[Stein \latin{et~al.}(2009)Stein, Kronik, and Baer]{Stein2009}
Stein,~T.; Kronik,~L.; Baer,~R. Reliable prediction of charge transfer
  excitations in molecular complexes using time-dependent density functional
  theory. \emph{J. Am. Chem. Soc.} \textbf{2009}, \emph{131}, 2818--2820\relax
\mciteBstWouldAddEndPuncttrue
\mciteSetBstMidEndSepPunct{\mcitedefaultmidpunct}
{\mcitedefaultendpunct}{\mcitedefaultseppunct}\relax
\EndOfBibitem
\bibitem[Kronik \latin{et~al.}(2012)Kronik, Stein, Refaely-Abramson, and
  Baer]{Kronik2012}
Kronik,~L.; Stein,~T.; Refaely-Abramson,~S.; Baer,~R. Excitation gaps of
  finite-sized systems from optimally tuned range-separated hybrid functionals.
  \emph{J. Chem. Theory and Comput.} \textbf{2012}, \emph{8}, 1515--1531\relax
\mciteBstWouldAddEndPuncttrue
\mciteSetBstMidEndSepPunct{\mcitedefaultmidpunct}
{\mcitedefaultendpunct}{\mcitedefaultseppunct}\relax
\EndOfBibitem
\bibitem[Körzdörfer and Brédas(2014)Körzdörfer, and
  Brédas]{Korzdorfer2014}
Körzdörfer,~T.; Brédas,~J.~L. Organic electronic materials: Recent advances
  in the DFT description of the ground and excited states using tuned
  range-separated hybrid functionals. \emph{Acc. Chem. Res.} \textbf{2014},
  \emph{47}, 3284--3291\relax
\mciteBstWouldAddEndPuncttrue
\mciteSetBstMidEndSepPunct{\mcitedefaultmidpunct}
{\mcitedefaultendpunct}{\mcitedefaultseppunct}\relax
\EndOfBibitem
\bibitem[Vandaele \latin{et~al.}(2022)Vandaele, Mali{\v{s}}, and
  Luber]{Vandaele2022}
Vandaele,~E.; Mali{\v{s}},~M.; Luber,~S. {The $\Delta$SCF method for
  non-adiabatic dynamics of systems in the liquid phase}. \emph{J. Chem. Phys}
  \textbf{2022}, \emph{156}, 130901\relax
\mciteBstWouldAddEndPuncttrue
\mciteSetBstMidEndSepPunct{\mcitedefaultmidpunct}
{\mcitedefaultendpunct}{\mcitedefaultseppunct}\relax
\EndOfBibitem
\bibitem[Levi \latin{et~al.}(2020)Levi, Ivanov, and J\'onsson]{Levi2020}
Levi,~G.; Ivanov,~A.~V.; J\'onsson,~H. Variational Density Functional
  Calculations of Excited States via Direct Optimization. \emph{J. Chem. Theory
  Comput.} \textbf{2020}, \emph{16}, 6968--6982\relax
\mciteBstWouldAddEndPuncttrue
\mciteSetBstMidEndSepPunct{\mcitedefaultmidpunct}
{\mcitedefaultendpunct}{\mcitedefaultseppunct}\relax
\EndOfBibitem
\bibitem[Carter-Fenk and Herbert(2020)Carter-Fenk, and
  Herbert]{Carter-Fenk2020}
Carter-Fenk,~K.; Herbert,~J.~M. State-Targeted Energy Projection: A Simple and
  Robust Approach to Orbital Relaxation of Non-Aufbau Self-Consistent Field
  Solutions. \emph{J. Chem. Theory Comput.} \textbf{2020}, \emph{16},
  5067--5082\relax
\mciteBstWouldAddEndPuncttrue
\mciteSetBstMidEndSepPunct{\mcitedefaultmidpunct}
{\mcitedefaultendpunct}{\mcitedefaultseppunct}\relax
\EndOfBibitem
\bibitem[Ayers \latin{et~al.}(2015)Ayers, Levy, and Nagy]{Ayers2015}
Ayers,~P.~W.; Levy,~M.; Nagy, Communication: Kohn-Sham theory for excited
  states of Coulomb systems. \emph{J. Chem. Phys.} \textbf{2015}, \emph{143},
  191101\relax
\mciteBstWouldAddEndPuncttrue
\mciteSetBstMidEndSepPunct{\mcitedefaultmidpunct}
{\mcitedefaultendpunct}{\mcitedefaultseppunct}\relax
\EndOfBibitem
\bibitem[Zhekova \latin{et~al.}(2014)Zhekova, Seth, and Ziegler]{Zhekova2014}
Zhekova,~H.~R.; Seth,~M.; Ziegler,~T. A perspective on the relative merits of
  time-dependent and time-independent density functional theory in studies of
  the electron spectra due to transition metal complexes. An illustration
  through applications to copper tetrachloride and plastocyanin. \emph{Int. J.
  Quantum Chem.} \textbf{2014}, \emph{114}, 1019--1029\relax
\mciteBstWouldAddEndPuncttrue
\mciteSetBstMidEndSepPunct{\mcitedefaultmidpunct}
{\mcitedefaultendpunct}{\mcitedefaultseppunct}\relax
\EndOfBibitem
\bibitem[Seidu \latin{et~al.}(2015)Seidu, Krykunov, and Ziegler]{Seidu2015}
Seidu,~I.; Krykunov,~M.; Ziegler,~T. Applications of time-dependent and
  time-independent density functional theory to Rydberg transitions. \emph{J.
  Phys. Chem.} \textbf{2015}, \emph{119}, 5107--5116\relax
\mciteBstWouldAddEndPuncttrue
\mciteSetBstMidEndSepPunct{\mcitedefaultmidpunct}
{\mcitedefaultendpunct}{\mcitedefaultseppunct}\relax
\EndOfBibitem
\bibitem[Cheng \latin{et~al.}(2008)Cheng, Wu, and Voorhis]{Cheng2008}
Cheng,~C.~L.; Wu,~Q.; Voorhis,~T.~V. Rydberg energies using excited state
  density functional theory. \emph{J. Chem. Phys.} \textbf{2008}, \emph{129},
  124112\relax
\mciteBstWouldAddEndPuncttrue
\mciteSetBstMidEndSepPunct{\mcitedefaultmidpunct}
{\mcitedefaultendpunct}{\mcitedefaultseppunct}\relax
\EndOfBibitem
\bibitem[Besley(2021)]{Besley2021}
Besley,~N.~A. Modeling of the spectroscopy of core electrons with density
  functional theory. \emph{Wiley Interdiscip. Rev. Comput. Mol. Sci.}
  \textbf{2021}, \emph{11}, e1527\relax
\mciteBstWouldAddEndPuncttrue
\mciteSetBstMidEndSepPunct{\mcitedefaultmidpunct}
{\mcitedefaultendpunct}{\mcitedefaultseppunct}\relax
\EndOfBibitem
\bibitem[Besley \latin{et~al.}(2009)Besley, Gilbert, and Gill]{Besley2009}
Besley,~N.~A.; Gilbert,~A.~T.; Gill,~P.~M. Self-consistent-field calculations
  of core excited states. \emph{J. Chem. Phys.} \textbf{2009}, \emph{130},
  124308\relax
\mciteBstWouldAddEndPuncttrue
\mciteSetBstMidEndSepPunct{\mcitedefaultmidpunct}
{\mcitedefaultendpunct}{\mcitedefaultseppunct}\relax
\EndOfBibitem
\bibitem[Hait and Head-Gordon(2020)Hait, and Head-Gordon]{Hait2020}
Hait,~D.; Head-Gordon,~M. Excited State Orbital Optimization via Minimizing the
  Square of the Gradient: General Approach and Application to Singly and Doubly
  Excited States via Density Functional Theory. \emph{J. Chem. Theory Comput.}
  \textbf{2020}, \emph{16}, 1699--1710\relax
\mciteBstWouldAddEndPuncttrue
\mciteSetBstMidEndSepPunct{\mcitedefaultmidpunct}
{\mcitedefaultendpunct}{\mcitedefaultseppunct}\relax
\EndOfBibitem
\bibitem[Vandaele \latin{et~al.}(2022)Vandaele, Mali{\v{s}}, and
  Luber]{Vandaele2022_2}
Vandaele,~E.; Mali{\v{s}},~M.; Luber,~S. {The photodissociation of solvated
  cyclopropanone and its hydrate explored via non-adiabatic molecular dynamics
  using $\Delta$SCF}. \emph{Phys. Chem. Chem. Phys.} \textbf{2022}, \emph{24},
  5669--5679\relax
\mciteBstWouldAddEndPuncttrue
\mciteSetBstMidEndSepPunct{\mcitedefaultmidpunct}
{\mcitedefaultendpunct}{\mcitedefaultseppunct}\relax
\EndOfBibitem
\bibitem[Mali{\v{s}} and Luber(2020)Mali{\v{s}}, and Luber]{Malis2020}
Mali{\v{s}},~M.; Luber,~S. {Trajectory Surface Hopping Nonadiabatic Molecular
  Dynamics with Kohn-Sham $\Delta$SCF for Condensed-Phase Systems}. \emph{J.
  Chem. Theory Comput} \textbf{2020}, \emph{16}, 4071--4086\relax
\mciteBstWouldAddEndPuncttrue
\mciteSetBstMidEndSepPunct{\mcitedefaultmidpunct}
{\mcitedefaultendpunct}{\mcitedefaultseppunct}\relax
\EndOfBibitem
\bibitem[Pradhan \latin{et~al.}(2018)Pradhan, Sato, and Akimov]{Pradhan2018}
Pradhan,~E.; Sato,~K.; Akimov,~A.~V. Non-adiabatic molecular dynamics with
  $\Delta$SCF excited states. \emph{J. Phys. Condens. Matter} \textbf{2018},
  \emph{30}, 484001\relax
\mciteBstWouldAddEndPuncttrue
\mciteSetBstMidEndSepPunct{\mcitedefaultmidpunct}
{\mcitedefaultendpunct}{\mcitedefaultseppunct}\relax
\EndOfBibitem
\bibitem[Levi \latin{et~al.}(2018)Levi, Pápai, Henriksen, Dohn, and
  Møller]{Levi2018}
Levi,~G.; Pápai,~M.; Henriksen,~N.~E.; Dohn,~A.~O.; Møller,~K.~B. Solution
  Structure and Ultrafast Vibrational Relaxation of the PtPOP Complex Revealed
  by $\Delta$SCF-QM/MM Direct Dynamics Simulations. \emph{J. Phys. Chem. C}
  \textbf{2018}, \emph{122}, 7100--7119\relax
\mciteBstWouldAddEndPuncttrue
\mciteSetBstMidEndSepPunct{\mcitedefaultmidpunct}
{\mcitedefaultendpunct}{\mcitedefaultseppunct}\relax
\EndOfBibitem
\bibitem[Burton(2022)]{Burton2022}
Burton,~H.~G. Energy Landscape of State-Specific Electronic Structure Theory.
  \emph{J. Chem. Theory Comput.} \textbf{2022}, \emph{18}, 1512--1526\relax
\mciteBstWouldAddEndPuncttrue
\mciteSetBstMidEndSepPunct{\mcitedefaultmidpunct}
{\mcitedefaultendpunct}{\mcitedefaultseppunct}\relax
\EndOfBibitem
\bibitem[Helgaker \latin{et~al.}(2014)Helgaker, Jørgensen, and
  Olsen]{MolecularElectronicStructureTheory}
Helgaker,~T.; Jørgensen,~P.; Olsen,~J. \emph{Molecular Electronic‐Structure
  Theory}; John Wiley \& Sons, Ltd, 2014; Chapter 4, pp 107--141\relax
\mciteBstWouldAddEndPuncttrue
\mciteSetBstMidEndSepPunct{\mcitedefaultmidpunct}
{\mcitedefaultendpunct}{\mcitedefaultseppunct}\relax
\EndOfBibitem
\bibitem[Jensen and J{\o}rgensen(1984)Jensen, and J{\o}rgensen]{Jensen1984}
Jensen,~H. J.~A.; J{\o}rgensen,~P. {A direct approach to second-order MCSCF
  calculations using a norm extended optimization scheme}. \emph{J. Chem.
  Phys.} \textbf{1984}, \emph{80}, 1204--1214\relax
\mciteBstWouldAddEndPuncttrue
\mciteSetBstMidEndSepPunct{\mcitedefaultmidpunct}
{\mcitedefaultendpunct}{\mcitedefaultseppunct}\relax
\EndOfBibitem
\bibitem[Olsen \latin{et~al.}(1983)Olsen, Yeager, and J{\o}rgensen]{Olsen1983}
Olsen,~J.; Yeager,~D.~L.; J{\o}rgensen,~P. {Optimization and Characterization
  of a Multiconfigurational Self-Consistent Field (MCSCF) State}. \emph{Adv.
  Chem. Phys.} \textbf{1983}, \emph{54}, 1--176\relax
\mciteBstWouldAddEndPuncttrue
\mciteSetBstMidEndSepPunct{\mcitedefaultmidpunct}
{\mcitedefaultendpunct}{\mcitedefaultseppunct}\relax
\EndOfBibitem
\bibitem[Golab \latin{et~al.}(1983)Golab, Yeager, and J{\o}rgensen]{Golab1983}
Golab,~J.~T.; Yeager,~D.~L.; J{\o}rgensen,~P. {Proper characterization of MC
  SCF stationary points}. \emph{Chem. Phys.} \textbf{1983}, \emph{78},
  175--199\relax
\mciteBstWouldAddEndPuncttrue
\mciteSetBstMidEndSepPunct{\mcitedefaultmidpunct}
{\mcitedefaultendpunct}{\mcitedefaultseppunct}\relax
\EndOfBibitem
\bibitem[Perdew and Levy(1985)Perdew, and Levy]{Perdew1985}
Perdew,~J.~P.; Levy,~M. Extrema of the density functional for the energy:
  Excited states from the ground-state theory. \emph{Phys. Rev. B, Condens.
  Matter} \textbf{1985}, \emph{31}, 6264\relax
\mciteBstWouldAddEndPuncttrue
\mciteSetBstMidEndSepPunct{\mcitedefaultmidpunct}
{\mcitedefaultendpunct}{\mcitedefaultseppunct}\relax
\EndOfBibitem
\bibitem[Kowalczyk \latin{et~al.}(2011)Kowalczyk, Yost, and
  Voorhis]{Kowalczyk2011}
Kowalczyk,~T.; Yost,~S.~R.; Voorhis,~T.~V. {Assessment of the $\Delta$SCF
  density functional theory approach for electronic excitations in organic
  dyes}. \emph{J. Chem. Phys.} \textbf{2011}, \emph{134}, 054128\relax
\mciteBstWouldAddEndPuncttrue
\mciteSetBstMidEndSepPunct{\mcitedefaultmidpunct}
{\mcitedefaultendpunct}{\mcitedefaultseppunct}\relax
\EndOfBibitem
\bibitem[Davidson(1975)]{Davidson1975}
Davidson,~E.~R. The iterative calculation of a few of the lowest eigenvalues
  and corresponding eigenvectors of large real-symmetric matrices. \emph{J.
  Comput. Phys.} \textbf{1975}, \emph{17}, 87--94\relax
\mciteBstWouldAddEndPuncttrue
\mciteSetBstMidEndSepPunct{\mcitedefaultmidpunct}
{\mcitedefaultendpunct}{\mcitedefaultseppunct}\relax
\EndOfBibitem
\bibitem[Pulay(1980)]{Pulay1980}
Pulay,~P. Convergence acceleration of iterative sequences. The case of SCF
  iteration. \emph{Chem. Phys. Lett.} \textbf{1980}, \emph{73}, 393--398\relax
\mciteBstWouldAddEndPuncttrue
\mciteSetBstMidEndSepPunct{\mcitedefaultmidpunct}
{\mcitedefaultendpunct}{\mcitedefaultseppunct}\relax
\EndOfBibitem
\bibitem[Pulay(1982)]{Pulay1982}
Pulay,~P. Improved SCF Convergence Acceleration. \emph{J. Comput. Chem.}
  \textbf{1982}, \emph{3}, 556--560\relax
\mciteBstWouldAddEndPuncttrue
\mciteSetBstMidEndSepPunct{\mcitedefaultmidpunct}
{\mcitedefaultendpunct}{\mcitedefaultseppunct}\relax
\EndOfBibitem
\bibitem[Taka \latin{et~al.}(2022)Taka, Lu, Gowland, Zuehlsdorff, Corzo,
  Pribram-Jones, Shi, Hratchian, and Isborn]{Taka2022}
Taka,~A.~A.; Lu,~S.-Y.; Gowland,~D.; Zuehlsdorff,~T.~J.; Corzo,~H.~H.;
  Pribram-Jones,~A.; Shi,~L.; Hratchian,~H.~P.; Isborn,~C.~M. {Comparison of
  Linear Response Theory, Projected Initial Maximum Overlap Method, and
  Molecular Dynamics-Based Vibronic Spectra: The Case of Methylene Blue}.
  \emph{J. Chem. Theory Comput.} \textbf{2022}, \emph{18}, 3039--3051\relax
\mciteBstWouldAddEndPuncttrue
\mciteSetBstMidEndSepPunct{\mcitedefaultmidpunct}
{\mcitedefaultendpunct}{\mcitedefaultseppunct}\relax
\EndOfBibitem
\bibitem[Gilbert \latin{et~al.}(2008)Gilbert, Besley, and Gill]{Gilbert2008}
Gilbert,~A.~T.; Besley,~N.~A.; Gill,~P.~M. Self-consistent field calculations
  of excited states using the maximum overlap method (MOM). \emph{J. Phys.
  Chem.} \textbf{2008}, \emph{112}, 13164--13171\relax
\mciteBstWouldAddEndPuncttrue
\mciteSetBstMidEndSepPunct{\mcitedefaultmidpunct}
{\mcitedefaultendpunct}{\mcitedefaultseppunct}\relax
\EndOfBibitem
\bibitem[Ivanov \latin{et~al.}(2021)Ivanov, Levi, J\'onsson, and
  J\'onsson]{Ivanov2021}
Ivanov,~A.~V.; Levi,~G.; J\'onsson,~E.~O.; J\'onsson,~H. Method for Calculating
  Excited Electronic States Using Density Functionals and Direct Orbital
  Optimization with Real Space Grid or Plane-Wave Basis Set. \emph{J. Chem.
  Theory Comput.} \textbf{2021}, \emph{17}, 5034--5049\relax
\mciteBstWouldAddEndPuncttrue
\mciteSetBstMidEndSepPunct{\mcitedefaultmidpunct}
{\mcitedefaultendpunct}{\mcitedefaultseppunct}\relax
\EndOfBibitem
\bibitem[Ivanov \latin{et~al.}(2021)Ivanov, J\'onsson, Vegge, and
  J\'onsson]{Ivanov2021_2}
Ivanov,~A.~V.; J\'onsson,~E.~O.; Vegge,~T.; J\'onsson,~H. Direct Energy
  Minimization Based on Exponential Transformation in Density Functional
  Calculations of Finite and Extended Systems. \emph{Comput. Phys. Commun.}
  \textbf{2021}, \emph{267}, 108047\relax
\mciteBstWouldAddEndPuncttrue
\mciteSetBstMidEndSepPunct{\mcitedefaultmidpunct}
{\mcitedefaultendpunct}{\mcitedefaultseppunct}\relax
\EndOfBibitem
\bibitem[Levi \latin{et~al.}(2020)Levi, Ivanov, and Jónsson]{Levi2020_2}
Levi,~G.; Ivanov,~A.~V.; Jónsson,~H. Variational calculations of excited
  states: Via direct optimization of the orbitals in DFT. \emph{Faraday
  Discuss.} \textbf{2020}, \emph{224}, 448--466\relax
\mciteBstWouldAddEndPuncttrue
\mciteSetBstMidEndSepPunct{\mcitedefaultmidpunct}
{\mcitedefaultendpunct}{\mcitedefaultseppunct}\relax
\EndOfBibitem
\bibitem[Voorhis and Head-Gordon(2002)Voorhis, and Head-Gordon]{VanVoorhis2002}
Voorhis,~T.~V.; Head-Gordon,~M. A geometric approach to direct minimization.
  \emph{Mol. Phys.} \textbf{2002}, \emph{100}, 1713--1721\relax
\mciteBstWouldAddEndPuncttrue
\mciteSetBstMidEndSepPunct{\mcitedefaultmidpunct}
{\mcitedefaultendpunct}{\mcitedefaultseppunct}\relax
\EndOfBibitem
\bibitem[Schmerwitz \latin{et~al.}(2022)Schmerwitz, Ivanov, J\'onsson,
  J\'onsson, and Levi]{Schmerwitz2022}
Schmerwitz,~Y. L.~A.; Ivanov,~A.~V.; J\'onsson,~E.~O.; J\'onsson,~H.; Levi,~G.
  Variational Density Functional Calculations of Excited States: Conical
  Intersection and Avoided Crossing in Ethylene Bond Twisting. \emph{J. Phys.
  Chem. Lett.} \textbf{2022}, \emph{13}, 3990--3999\relax
\mciteBstWouldAddEndPuncttrue
\mciteSetBstMidEndSepPunct{\mcitedefaultmidpunct}
{\mcitedefaultendpunct}{\mcitedefaultseppunct}\relax
\EndOfBibitem
\bibitem[Perdew \latin{et~al.}(1996)Perdew, Burke, and Ernzerhof]{Perdew1996}
Perdew,~J.~P.; Burke,~K.; Ernzerhof,~M. Generalized Gradient Approximation Made
  Simple. \emph{Phys. Rev. Lett.} \textbf{1996}, \emph{77}, 3865\relax
\mciteBstWouldAddEndPuncttrue
\mciteSetBstMidEndSepPunct{\mcitedefaultmidpunct}
{\mcitedefaultendpunct}{\mcitedefaultseppunct}\relax
\EndOfBibitem
\bibitem[Perdew \latin{et~al.}(1997)Perdew, Burke, and Ernzerhof]{Perdew1997}
Perdew,~J.~P.; Burke,~K.; Ernzerhof,~M. Erratum: Generalized Gradient
  Approximation Made Simple. \emph{Phys. Rev. Lett.} \textbf{1997}, \emph{78},
  1396\relax
\mciteBstWouldAddEndPuncttrue
\mciteSetBstMidEndSepPunct{\mcitedefaultmidpunct}
{\mcitedefaultendpunct}{\mcitedefaultseppunct}\relax
\EndOfBibitem
\bibitem[Jake \latin{et~al.}(2018)Jake, Henderson, and Scuseria]{Jake2018}
Jake,~L.~C.; Henderson,~T.~M.; Scuseria,~G.~E. Hartree-Fock symmetry breaking
  around conical intersections. \emph{J. Chem. Phys.} \textbf{2018},
  \emph{148}, 024109\relax
\mciteBstWouldAddEndPuncttrue
\mciteSetBstMidEndSepPunct{\mcitedefaultmidpunct}
{\mcitedefaultendpunct}{\mcitedefaultseppunct}\relax
\EndOfBibitem
\bibitem[Tóth and Pulay(2016)Tóth, and Pulay]{Toth2016}
Tóth,~Z.; Pulay,~P. Finding symmetry breaking Hartree-Fock solutions: The case
  of triplet instability. \emph{J. Chem. Phys.} \textbf{2016}, \emph{145},
  164102\relax
\mciteBstWouldAddEndPuncttrue
\mciteSetBstMidEndSepPunct{\mcitedefaultmidpunct}
{\mcitedefaultendpunct}{\mcitedefaultseppunct}\relax
\EndOfBibitem
\bibitem[Jiménez-Hoyos \latin{et~al.}(2011)Jiménez-Hoyos, Henderson, and
  Scuseria]{Jimenez-Hoyos2011}
Jiménez-Hoyos,~C.~A.; Henderson,~T.~M.; Scuseria,~G.~E. Generalized
  Hartree-Fock description of molecular dissociation. \emph{J. Chem. Theory
  Comput.} \textbf{2011}, \emph{7}, 2667--2674\relax
\mciteBstWouldAddEndPuncttrue
\mciteSetBstMidEndSepPunct{\mcitedefaultmidpunct}
{\mcitedefaultendpunct}{\mcitedefaultseppunct}\relax
\EndOfBibitem
\bibitem[Li and Paldus(2009)Li, and Paldus]{Li2009}
Li,~X.; Paldus,~J. Do independent-particle-model broken-symmetry solutions
  contain more physics than the symmetry-adapted ones? The case of homonuclear
  diatomics. \emph{J. Chem. Phys.} \textbf{2009}, \emph{130}, 084110\relax
\mciteBstWouldAddEndPuncttrue
\mciteSetBstMidEndSepPunct{\mcitedefaultmidpunct}
{\mcitedefaultendpunct}{\mcitedefaultseppunct}\relax
\EndOfBibitem
\bibitem[Coulson and Fischer(1949)Coulson, and Fischer]{Coulson1949}
Coulson,~C.~A.; Fischer,~I. XXXIV. Notes on the molecular orbital treatment of
  the hydrogen molecule. \emph{London Edinburgh Philos. Mag. J. Sci.}
  \textbf{1949}, \emph{40}, 386--393\relax
\mciteBstWouldAddEndPuncttrue
\mciteSetBstMidEndSepPunct{\mcitedefaultmidpunct}
{\mcitedefaultendpunct}{\mcitedefaultseppunct}\relax
\EndOfBibitem
\bibitem[Perdew \latin{et~al.}(2021)Perdew, Ruzsinszky, Sun, Nepal, and
  Kaplan]{Perdew2021}
Perdew,~J.~P.; Ruzsinszky,~A.; Sun,~J.; Nepal,~N.~K.; Kaplan,~A.~D.
  Interpretations of ground-state symmetry breaking and strong correlation in
  wavefunction and density functional theories. \emph{Proc. Natl. Acad. Sci.}
  \textbf{2021}, \emph{118}, e2017850118\relax
\mciteBstWouldAddEndPuncttrue
\mciteSetBstMidEndSepPunct{\mcitedefaultmidpunct}
{\mcitedefaultendpunct}{\mcitedefaultseppunct}\relax
\EndOfBibitem
\bibitem[Yu \latin{et~al.}(2016)Yu, Li, and Truhlar]{Yu2016}
Yu,~H.~S.; Li,~S.~L.; Truhlar,~D.~G. Perspective: Kohn-Sham density functional
  theory descending a staircase. \emph{J. Chem. Phys.} \textbf{2016},
  \emph{145}, 130901\relax
\mciteBstWouldAddEndPuncttrue
\mciteSetBstMidEndSepPunct{\mcitedefaultmidpunct}
{\mcitedefaultendpunct}{\mcitedefaultseppunct}\relax
\EndOfBibitem
\bibitem[Cohen \latin{et~al.}(2008)Cohen, Mori-Sánchez, and Yang]{Cohen2008}
Cohen,~A.~J.; Mori-Sánchez,~P.; Yang,~W. Insights into current limitations of
  density functional theory. \emph{Science} \textbf{2008}, \emph{321},
  792--794\relax
\mciteBstWouldAddEndPuncttrue
\mciteSetBstMidEndSepPunct{\mcitedefaultmidpunct}
{\mcitedefaultendpunct}{\mcitedefaultseppunct}\relax
\EndOfBibitem
\bibitem[Gräfenstein \latin{et~al.}(2002)Gräfenstein, Kraka, Filatov, and
  Cremer]{Grafenstein2002}
Gräfenstein,~J.; Kraka,~E.; Filatov,~M.; Cremer,~D. Can Unrestricted
  Density-Functional Theory Describe Open Shell Singlet Biradicals? \emph{Int.
  J. Mol. Sci} \textbf{2002}, \emph{3}, 360--394\relax
\mciteBstWouldAddEndPuncttrue
\mciteSetBstMidEndSepPunct{\mcitedefaultmidpunct}
{\mcitedefaultendpunct}{\mcitedefaultseppunct}\relax
\EndOfBibitem
\bibitem[Cremer \latin{et~al.}(2002)Cremer, Filatov, Polo, Kraka, and
  Shaik]{Cremer2002}
Cremer,~D.; Filatov,~M.; Polo,~V.; Kraka,~E.; Shaik,~S. Implicit and Explicit
  Coverage of Multi-reference Effects by Density Functional Theory. \emph{Int.
  J. Mol. Sci} \textbf{2002}, \emph{3}, 604--638\relax
\mciteBstWouldAddEndPuncttrue
\mciteSetBstMidEndSepPunct{\mcitedefaultmidpunct}
{\mcitedefaultendpunct}{\mcitedefaultseppunct}\relax
\EndOfBibitem
\bibitem[Cremer(2001)]{Cremer2001}
Cremer,~D. Density functional theory: Coverage of dynamic and non-dynamic
  electron correlation effects. \emph{Mol. Phys.} \textbf{2001}, \emph{99},
  1899--1940\relax
\mciteBstWouldAddEndPuncttrue
\mciteSetBstMidEndSepPunct{\mcitedefaultmidpunct}
{\mcitedefaultendpunct}{\mcitedefaultseppunct}\relax
\EndOfBibitem
\bibitem[Gräfenstein \latin{et~al.}(2000)Gräfenstein, Hjerpe, Kraka, and
  Cremer]{Grafenstein2000}
Gräfenstein,~J.; Hjerpe,~A.~M.; Kraka,~E.; Cremer,~D. An accurate description
  of the Bergman reaction using restricted and unrestricted DFT: Stability
  test, spin density, and on-top pair density. \emph{J. Phys. Chem.}
  \textbf{2000}, \emph{104}, 1748--1761\relax
\mciteBstWouldAddEndPuncttrue
\mciteSetBstMidEndSepPunct{\mcitedefaultmidpunct}
{\mcitedefaultendpunct}{\mcitedefaultseppunct}\relax
\EndOfBibitem
\bibitem[Wittbrodt and Schlegel(1996)Wittbrodt, and Schlegel]{Wittbrodt1996}
Wittbrodt,~J.~M.; Schlegel,~H.~B. Some reasons not to use spin projected
  density functional theory. \emph{J. Chem. Phys.} \textbf{1996}, \emph{105},
  6574--6577\relax
\mciteBstWouldAddEndPuncttrue
\mciteSetBstMidEndSepPunct{\mcitedefaultmidpunct}
{\mcitedefaultendpunct}{\mcitedefaultseppunct}\relax
\EndOfBibitem
\bibitem[Mali{\v{s}} \latin{et~al.}(2022)Mali{\v{s}}, Vandaele, and
  Luber]{Malis2022}
Mali{\v{s}},~M.; Vandaele,~E.; Luber,~S. {Spin-Orbit Couplings for Nonadiabatic
  Molecular Dynamics at the $\Delta$SCF Level}. \emph{J. Chem. Theory Comput.}
  \textbf{2022}, \emph{18}, 4082--4094\relax
\mciteBstWouldAddEndPuncttrue
\mciteSetBstMidEndSepPunct{\mcitedefaultmidpunct}
{\mcitedefaultendpunct}{\mcitedefaultseppunct}\relax
\EndOfBibitem
\bibitem[Vaucher and Reiher(2017)Vaucher, and Reiher]{Vaucher2017}
Vaucher,~A.~C.; Reiher,~M. Steering Orbital Optimization out of Local Minima
  and Saddle Points Toward Lower Energy. \emph{J. Chem. Theory Comput.}
  \textbf{2017}, \emph{13}, 1219--1228\relax
\mciteBstWouldAddEndPuncttrue
\mciteSetBstMidEndSepPunct{\mcitedefaultmidpunct}
{\mcitedefaultendpunct}{\mcitedefaultseppunct}\relax
\EndOfBibitem
\bibitem[Pelzer and Wigner(1932)Pelzer, and Wigner]{Pelzer1932}
Pelzer,~H.; Wigner,~E.~P. Über die Geschwindigkeitskonstante von
  Austauschreaktionen. \emph{Z. Phys. Chem. B} \textbf{1932}, \emph{15},
  445--471\relax
\mciteBstWouldAddEndPuncttrue
\mciteSetBstMidEndSepPunct{\mcitedefaultmidpunct}
{\mcitedefaultendpunct}{\mcitedefaultseppunct}\relax
\EndOfBibitem
\bibitem[Eyring(1935)]{Eyring1935}
Eyring,~H. The activated complex in chemical reactions. \emph{J. Chem. Phys.}
  \textbf{1935}, \emph{3}, 63--71\relax
\mciteBstWouldAddEndPuncttrue
\mciteSetBstMidEndSepPunct{\mcitedefaultmidpunct}
{\mcitedefaultendpunct}{\mcitedefaultseppunct}\relax
\EndOfBibitem
\bibitem[Wigner(1938)]{Wigner1938}
Wigner,~E. The transition state method. \emph{Trans. Faraday Soc.}
  \textbf{1938}, \emph{34}, 29--41\relax
\mciteBstWouldAddEndPuncttrue
\mciteSetBstMidEndSepPunct{\mcitedefaultmidpunct}
{\mcitedefaultendpunct}{\mcitedefaultseppunct}\relax
\EndOfBibitem
\bibitem[Cerjan and Miller(1981)Cerjan, and Miller]{Cerjan1981}
Cerjan,~C.~J.; Miller,~W.~H. On finding transition states. \emph{J. Chem.
  Phys.} \textbf{1981}, \emph{75}, 2800--2801\relax
\mciteBstWouldAddEndPuncttrue
\mciteSetBstMidEndSepPunct{\mcitedefaultmidpunct}
{\mcitedefaultendpunct}{\mcitedefaultseppunct}\relax
\EndOfBibitem
\bibitem[Simons \latin{et~al.}(1983)Simons, Jorgensen, Taylor, and
  Orment]{Simons1983}
Simons,~J.; Jorgensen,~P.; Taylor,~H.; Orment,~J. Walking on Potential Energy
  Surfaces. \emph{J. Phys. Chem} \textbf{1983}, \emph{87}, 2745--2753\relax
\mciteBstWouldAddEndPuncttrue
\mciteSetBstMidEndSepPunct{\mcitedefaultmidpunct}
{\mcitedefaultendpunct}{\mcitedefaultseppunct}\relax
\EndOfBibitem
\bibitem[Banerjee \latin{et~al.}(1985)Banerjee, Adams, Simons, and
  Shepard]{Banerjee1985}
Banerjee,~A.; Adams,~N.; Simons,~J.; Shepard,~R. Search for Stationary Points
  on Surfaces. \emph{J. Phys. Chem.} \textbf{1985}, \emph{89}, 52--57\relax
\mciteBstWouldAddEndPuncttrue
\mciteSetBstMidEndSepPunct{\mcitedefaultmidpunct}
{\mcitedefaultendpunct}{\mcitedefaultseppunct}\relax
\EndOfBibitem
\bibitem[Marie and Burton(2023)Marie, and Burton]{Marie2013}
Marie,~A.; Burton,~H. G.~A. Excited states, symmetry breaking, and unphysical
  solutions in state-specific CASSCF theory. 2023;
  \url{https://arxiv.org/abs/2301.11731}\relax
\mciteBstWouldAddEndPuncttrue
\mciteSetBstMidEndSepPunct{\mcitedefaultmidpunct}
{\mcitedefaultendpunct}{\mcitedefaultseppunct}\relax
\EndOfBibitem
\bibitem[Hoffmann \latin{et~al.}(2002)Hoffmann, Sherrill, Leininger, and
  Schaefer]{Hoffmann2002}
Hoffmann,~M.~R.; Sherrill,~C.~D.; Leininger,~M.~L.; Schaefer,~H.~F.
  {Optimization of MCSCF excited states using directions of negative
  curvature}. \emph{Chem. Phys. Lett.} \textbf{2002}, \emph{355},
  183--192\relax
\mciteBstWouldAddEndPuncttrue
\mciteSetBstMidEndSepPunct{\mcitedefaultmidpunct}
{\mcitedefaultendpunct}{\mcitedefaultseppunct}\relax
\EndOfBibitem
\bibitem[Murtagh and Sargent(1970)Murtagh, and Sargent]{Murtagh1970}
Murtagh,~B.~A.; Sargent,~R. W.~H. Computational experience with quadratically
  convergent minimisation methods. \emph{Comput. J.} \textbf{1970}, \emph{13},
  185--194\relax
\mciteBstWouldAddEndPuncttrue
\mciteSetBstMidEndSepPunct{\mcitedefaultmidpunct}
{\mcitedefaultendpunct}{\mcitedefaultseppunct}\relax
\EndOfBibitem
\bibitem[Powell(1973)]{Powell1973}
Powell,~M. J.~D. A New Algorithm for Unconstrained Optimization.
  \emph{Nonlinear Programming} \textbf{1973}, 31--65\relax
\mciteBstWouldAddEndPuncttrue
\mciteSetBstMidEndSepPunct{\mcitedefaultmidpunct}
{\mcitedefaultendpunct}{\mcitedefaultseppunct}\relax
\EndOfBibitem
\bibitem[Bofill(1994)]{Bofill1994}
Bofill,~J.~M. Updated Hessian Matrix and the Restricted Step Method for
  Locating Transition Structures. \emph{J. Comput. Chem.} \textbf{1994},
  \emph{15}, 1--11\relax
\mciteBstWouldAddEndPuncttrue
\mciteSetBstMidEndSepPunct{\mcitedefaultmidpunct}
{\mcitedefaultendpunct}{\mcitedefaultseppunct}\relax
\EndOfBibitem
\bibitem[Henkelman and Jónsson(1999)Henkelman, and Jónsson]{Henkelman1999}
Henkelman,~G.; Jónsson,~H. A dimer method for finding saddle points on high
  dimensional potential surfaces using only first derivatives. \emph{J. Chem.
  Phys.} \textbf{1999}, \emph{111}, 7010--7022\relax
\mciteBstWouldAddEndPuncttrue
\mciteSetBstMidEndSepPunct{\mcitedefaultmidpunct}
{\mcitedefaultendpunct}{\mcitedefaultseppunct}\relax
\EndOfBibitem
\bibitem[Olsen \latin{et~al.}(2004)Olsen, Kroes, Henkelman, Arnaldsson, and
  Jónsson]{Olsen2004}
Olsen,~R.~A.; Kroes,~G.~J.; Henkelman,~G.; Arnaldsson,~A.; Jónsson,~H.
  Comparison of methods for finding saddle points without knowledge of the
  final states. \emph{J. Chem. Phys.} \textbf{2004}, \emph{121},
  9776--9792\relax
\mciteBstWouldAddEndPuncttrue
\mciteSetBstMidEndSepPunct{\mcitedefaultmidpunct}
{\mcitedefaultendpunct}{\mcitedefaultseppunct}\relax
\EndOfBibitem
\bibitem[Kästner and Sherwood(2008)Kästner, and Sherwood]{Kastner2008}
Kästner,~J.; Sherwood,~P. Superlinearly converging dimer method for transition
  state search. \emph{J. Chem. Phys.} \textbf{2008}, \emph{128}, 014106\relax
\mciteBstWouldAddEndPuncttrue
\mciteSetBstMidEndSepPunct{\mcitedefaultmidpunct}
{\mcitedefaultendpunct}{\mcitedefaultseppunct}\relax
\EndOfBibitem
\bibitem[Manuel Plasencia~Gutiérrez and Jónsson(2016)Manuel
  Plasencia~Gutiérrez, and Jónsson]{Gutierrez2016}
Manuel Plasencia~Gutiérrez,~C.~A.; Jónsson,~H. Improved Minimum Mode
  Following Method for Finding First Order Saddle Points. \emph{J. Chem. Theory
  Comput.} \textbf{2016}, \emph{13}, 125--134\relax
\mciteBstWouldAddEndPuncttrue
\mciteSetBstMidEndSepPunct{\mcitedefaultmidpunct}
{\mcitedefaultendpunct}{\mcitedefaultseppunct}\relax
\EndOfBibitem
\bibitem[Lánczos(1950)]{Lanczos1950}
Lánczos,~C. An Iteration Method for the Solution of the Eigenvalue Problem of
  Linear Differential and Integral Operators. \emph{J. Res. Natl. Bur. Stand.}
  \textbf{1950}, \emph{45}, 255--282\relax
\mciteBstWouldAddEndPuncttrue
\mciteSetBstMidEndSepPunct{\mcitedefaultmidpunct}
{\mcitedefaultendpunct}{\mcitedefaultseppunct}\relax
\EndOfBibitem
\bibitem[Crouzeix \latin{et~al.}(1994)Crouzeix, Philippe, and
  Sadkane]{Crouzeix1994}
Crouzeix,~M.; Philippe,~B.; Sadkane,~M. The Davidson Method. \emph{SIAM J. Sci.
  Comput.} \textbf{1994}, \emph{15}, 62--76\relax
\mciteBstWouldAddEndPuncttrue
\mciteSetBstMidEndSepPunct{\mcitedefaultmidpunct}
{\mcitedefaultendpunct}{\mcitedefaultseppunct}\relax
\EndOfBibitem
\bibitem[Lehtola \latin{et~al.}(2020)Lehtola, Blockhuys, and {Van
  Alsenoy}]{Lehtola2020}
Lehtola,~S.; Blockhuys,~F.; {Van Alsenoy},~C. {An overview of self-consistent
  field calculations within finite basis sets}. \emph{Molecules} \textbf{2020},
  \emph{25}, 1--23\relax
\mciteBstWouldAddEndPuncttrue
\mciteSetBstMidEndSepPunct{\mcitedefaultmidpunct}
{\mcitedefaultendpunct}{\mcitedefaultseppunct}\relax
\EndOfBibitem
\bibitem[Perdew and Zunger(1981)Perdew, and Zunger]{Perdew1981}
Perdew,~J.~P.; Zunger,~A. Self-interaction correction to density-functional
  approximations for many-electron systems. \emph{Phys. Rev. B, Condens.
  Matter} \textbf{1981}, \emph{23}, 5048--5079\relax
\mciteBstWouldAddEndPuncttrue
\mciteSetBstMidEndSepPunct{\mcitedefaultmidpunct}
{\mcitedefaultendpunct}{\mcitedefaultseppunct}\relax
\EndOfBibitem
\bibitem[Lehtola \latin{et~al.}(2016)Lehtola, Head-Gordon, and
  Jónsson]{Lehtola2016}
Lehtola,~S.; Head-Gordon,~M.; Jónsson,~H. {Complex orbitals, multiple local
  minima and symmetry breaking in Perdew-Zunger self-interaction corrected
  density-functional theory calculations}. \emph{J. Chem. Theory Comput.}
  \textbf{2016}, \emph{12}, 3195\relax
\mciteBstWouldAddEndPuncttrue
\mciteSetBstMidEndSepPunct{\mcitedefaultmidpunct}
{\mcitedefaultendpunct}{\mcitedefaultseppunct}\relax
\EndOfBibitem
\bibitem[Baker(1905)]{Baker1905}
Baker,~H.~F. Alternants and Continuous Groups. \emph{Proc. London Math. Soc.}
  \textbf{1905}, \emph{2}, 24--47\relax
\mciteBstWouldAddEndPuncttrue
\mciteSetBstMidEndSepPunct{\mcitedefaultmidpunct}
{\mcitedefaultendpunct}{\mcitedefaultseppunct}\relax
\EndOfBibitem
\bibitem[Campbell(1896)]{Campbell1896}
Campbell,~J.~E. On a Law of Combination of Operators bearing on the Theory of
  Continuous Transformation Groups. \emph{Proc. London Math. Soc.}
  \textbf{1896}, \emph{s1-28}, 381--390\relax
\mciteBstWouldAddEndPuncttrue
\mciteSetBstMidEndSepPunct{\mcitedefaultmidpunct}
{\mcitedefaultendpunct}{\mcitedefaultseppunct}\relax
\EndOfBibitem
\bibitem[Campbell(1897)]{Campbell1897}
Campbell,~J.~E. On a Law of Combination of Operators (Second Paper).
  \emph{Proc. London Math. Soc.} \textbf{1897}, \emph{s1-29}, 14--32\relax
\mciteBstWouldAddEndPuncttrue
\mciteSetBstMidEndSepPunct{\mcitedefaultmidpunct}
{\mcitedefaultendpunct}{\mcitedefaultseppunct}\relax
\EndOfBibitem
\bibitem[Hausdorff(1906)]{Hausdorff1906}
Hausdorff,~F. Die symbolische Exponentialformel in der Gruppentheorie.
  \emph{Ber. Verh. Kgl. S\"achs. Ges. Wiss. Leipzig., Math.-phys. Kl.}
  \textbf{1906}, \emph{58}, 19--48\relax
\mciteBstWouldAddEndPuncttrue
\mciteSetBstMidEndSepPunct{\mcitedefaultmidpunct}
{\mcitedefaultendpunct}{\mcitedefaultseppunct}\relax
\EndOfBibitem
\bibitem[Salem and Rowland(1972)Salem, and Rowland]{Salem1972}
Salem,~L.; Rowland,~C. {The electronic properties of diradicals}. \emph{Angew.
  Chem. Int. Ed.} \textbf{1972}, \emph{11}, 92--111\relax
\mciteBstWouldAddEndPuncttrue
\mciteSetBstMidEndSepPunct{\mcitedefaultmidpunct}
{\mcitedefaultendpunct}{\mcitedefaultseppunct}\relax
\EndOfBibitem
\bibitem[Gould \latin{et~al.}(2014)Gould, Ortner, and Packwood]{Gould2014}
Gould,~N.; Ortner,~C.; Packwood,~D. An Efficient Dimer Method With
  Preconditioning And Linesearch. 2014;
  \url{http://arxiv.org/abs/1407.2817}\relax
\mciteBstWouldAddEndPuncttrue
\mciteSetBstMidEndSepPunct{\mcitedefaultmidpunct}
{\mcitedefaultendpunct}{\mcitedefaultseppunct}\relax
\EndOfBibitem
\bibitem[Sharada \latin{et~al.}(2015)Sharada, Stück, Sundstrom, Bell, and
  Head-Gordon]{Sharada2015}
Sharada,~S.~M.; Stück,~D.; Sundstrom,~E.~J.; Bell,~A.~T.; Head-Gordon,~M.
  Wavefunction stability analysis without analytical electronic Hessians:
  Application to orbital-optimised second-order Møller-Plesset theory and
  VV10-containing density functionals. \emph{Mol. Phys.} \textbf{2015},
  \emph{113}, 1802--1808\relax
\mciteBstWouldAddEndPuncttrue
\mciteSetBstMidEndSepPunct{\mcitedefaultmidpunct}
{\mcitedefaultendpunct}{\mcitedefaultseppunct}\relax
\EndOfBibitem
\bibitem[Mortensen \latin{et~al.}(2005)Mortensen, Hansen, and
  Jacobsen]{Mortensen2005}
Mortensen,~J.~J.; Hansen,~L.~B.; Jacobsen,~K.~W. Real-space grid implementation
  of the projector augmented wave method. \emph{Phys. Rev. B, Condens. Matter}
  \textbf{2005}, \emph{71}, 035109\relax
\mciteBstWouldAddEndPuncttrue
\mciteSetBstMidEndSepPunct{\mcitedefaultmidpunct}
{\mcitedefaultendpunct}{\mcitedefaultseppunct}\relax
\EndOfBibitem
\bibitem[Enkovaara \latin{et~al.}(2010)Enkovaara, Rostgaard, Mortensen, Chen,
  Dułak, Ferrighi, Gavnholt, Glinsvad, Haikola, Hansen, Kristoffersen, Kuisma,
  Larsen, Lehtovaara, Ljungberg, Lopez-Acevedo, Moses, Ojanen, Olsen, Petzold,
  Romero, Stausholm-Møller, Strange, Tritsaris, Vanin, Walter, Hammer,
  Häkkinen, Madsen, Nieminen, Nørskov, Puska, Rantala, Schiøtz, Thygesen,
  and Jacobsen]{Enkovaara2010}
Enkovaara,~J.; Rostgaard,~C.; Mortensen,~J.~J.; Chen,~J.; Dułak,~M.;
  Ferrighi,~L.; Gavnholt,~J.; Glinsvad,~C.; Haikola,~V.; Hansen,~H.~A.;
  Kristoffersen,~H.~H.; Kuisma,~M.; Larsen,~A.~H.; Lehtovaara,~L.;
  Ljungberg,~M.; Lopez-Acevedo,~O.; Moses,~P.~G.; Ojanen,~J.; Olsen,~T.;
  Petzold,~V.; Romero,~N.~A.; Stausholm-Møller,~J.; Strange,~M.;
  Tritsaris,~G.~A.; Vanin,~M.; Walter,~M.; Hammer,~B.; Häkkinen,~H.;
  Madsen,~G.~K.; Nieminen,~R.~M.; Nørskov,~J.~K.; Puska,~M.; Rantala,~T.~T.;
  Schiøtz,~J.; Thygesen,~K.~S.; Jacobsen,~K.~W. Electronic structure
  calculations with GPAW: A real-space implementation of the projector
  augmented-wave method. \emph{J. Phys. Condens. Matter} \textbf{2010},
  \emph{22}, 253202\relax
\mciteBstWouldAddEndPuncttrue
\mciteSetBstMidEndSepPunct{\mcitedefaultmidpunct}
{\mcitedefaultendpunct}{\mcitedefaultseppunct}\relax
\EndOfBibitem
\bibitem[Larsen \latin{et~al.}(2009)Larsen, Vanin, Mortensen, Thygesen, and
  Jacobsen]{Larsen2009}
Larsen,~A.~H.; Vanin,~M.; Mortensen,~J.~J.; Thygesen,~K.~S.; Jacobsen,~K.~W.
  Localized atomic basis set in the projector augmented wave method.
  \emph{Phys. Rev. B, Condens. Matter} \textbf{2009}, \emph{80}, 195112\relax
\mciteBstWouldAddEndPuncttrue
\mciteSetBstMidEndSepPunct{\mcitedefaultmidpunct}
{\mcitedefaultendpunct}{\mcitedefaultseppunct}\relax
\EndOfBibitem
\bibitem[Dunning(1989)]{Dunning1989}
Dunning,~T.~H. Gaussian basis sets for use in correlated molecular
  calculations. I. The atoms boron through neon and hydrogen. \emph{J. Chem.
  Phys.} \textbf{1989}, \emph{90}, 1007--1023\relax
\mciteBstWouldAddEndPuncttrue
\mciteSetBstMidEndSepPunct{\mcitedefaultmidpunct}
{\mcitedefaultendpunct}{\mcitedefaultseppunct}\relax
\EndOfBibitem
\bibitem[Kendall \latin{et~al.}(1992)Kendall, Dunning, and
  Harrison]{Kendall1992}
Kendall,~R.~A.; Dunning,~T.~H.; Harrison,~R.~J. Electron affinities of the
  first-row atoms revisited. Systematic basis sets and wave functions. \emph{J.
  Chem. Phys.} \textbf{1992}, \emph{96}, 6796--6806\relax
\mciteBstWouldAddEndPuncttrue
\mciteSetBstMidEndSepPunct{\mcitedefaultmidpunct}
{\mcitedefaultendpunct}{\mcitedefaultseppunct}\relax
\EndOfBibitem
\bibitem[Woon and Dunning(1994)Woon, and Dunning]{Woon1994}
Woon,~D.~E.; Dunning,~T.~H. Gaussian basis sets for use in correlated molecular
  calculations. IV. Calculation of static electrical response properties.
  \emph{J. Chem. Phys.} \textbf{1994}, \emph{100}, 2975--2988\relax
\mciteBstWouldAddEndPuncttrue
\mciteSetBstMidEndSepPunct{\mcitedefaultmidpunct}
{\mcitedefaultendpunct}{\mcitedefaultseppunct}\relax
\EndOfBibitem
\bibitem[Weigend and Ahlrichs(2005)Weigend, and Ahlrichs]{Weigend2005}
Weigend,~F.; Ahlrichs,~R. Balanced basis sets of split valence, triple zeta
  valence and quadruple zeta valence quality for H to Rn: Design and assessment
  of accuracy. \emph{Phys. Chem. Chem. Phys.} \textbf{2005}, \emph{7},
  3297--3305\relax
\mciteBstWouldAddEndPuncttrue
\mciteSetBstMidEndSepPunct{\mcitedefaultmidpunct}
{\mcitedefaultendpunct}{\mcitedefaultseppunct}\relax
\EndOfBibitem
\bibitem[Rossi \latin{et~al.}(2015)Rossi, Lehtola, Sakko, Puska, and
  Nieminen]{Rossi2015}
Rossi,~T.~P.; Lehtola,~S.; Sakko,~A.; Puska,~M.~J.; Nieminen,~R.~M.
  Nanoplasmonics simulations at the basis set limit through
  completeness-optimized, local numerical basis sets. \emph{J. Chem. Phys.}
  \textbf{2015}, \emph{142}, 094114\relax
\mciteBstWouldAddEndPuncttrue
\mciteSetBstMidEndSepPunct{\mcitedefaultmidpunct}
{\mcitedefaultendpunct}{\mcitedefaultseppunct}\relax
\EndOfBibitem
\bibitem[Bl\"ochl(1994)]{Blochl1994}
Bl\"ochl,~P.~E. Projector augmented-wave method. \emph{Phys. Rev. B, Condens.
  Matter} \textbf{1994}, \emph{50}, 17953\relax
\mciteBstWouldAddEndPuncttrue
\mciteSetBstMidEndSepPunct{\mcitedefaultmidpunct}
{\mcitedefaultendpunct}{\mcitedefaultseppunct}\relax
\EndOfBibitem
\bibitem[Loos \latin{et~al.}(2021)Loos, Comin, Blase, and Jacquemin]{Loos2021}
Loos,~P.~F.; Comin,~M.; Blase,~X.; Jacquemin,~D. Reference Energies for
  Intramolecular Charge-Transfer Excitations. \emph{J. Chem. Theory Comput.}
  \textbf{2021}, \emph{17}, 3666--3686\relax
\mciteBstWouldAddEndPuncttrue
\mciteSetBstMidEndSepPunct{\mcitedefaultmidpunct}
{\mcitedefaultendpunct}{\mcitedefaultseppunct}\relax
\EndOfBibitem
\bibitem[Lehtola \latin{et~al.}(2018)Lehtola, Steigemann, Oliveira, and
  Marques]{Lehtola2018}
Lehtola,~S.; Steigemann,~C.; Oliveira,~M.~J.; Marques,~M.~A. Recent
  developments in LIBXC — A comprehensive library of functionals for density
  functional theory. \emph{SoftwareX} \textbf{2018}, \emph{7}, 1--5\relax
\mciteBstWouldAddEndPuncttrue
\mciteSetBstMidEndSepPunct{\mcitedefaultmidpunct}
{\mcitedefaultendpunct}{\mcitedefaultseppunct}\relax
\EndOfBibitem
\bibitem[Barbatti and Crespo-Otero(2016)Barbatti, and
  Crespo-Otero]{Barbatti2014}
Barbatti,~M.; Crespo-Otero,~R. In \emph{Density-Functional Methods for Excited
  States}; Ferr{\'e},~N., Filatov,~M., Huix-Rotllant,~M., Eds.; Springer
  International Publishing: Cham, 2016; pp 415--444\relax
\mciteBstWouldAddEndPuncttrue
\mciteSetBstMidEndSepPunct{\mcitedefaultmidpunct}
{\mcitedefaultendpunct}{\mcitedefaultseppunct}\relax
\EndOfBibitem
\bibitem[Mewes \latin{et~al.}(2014)Mewes, Jovanović, Marian, and
  Dreuw]{Mewes2014}
Mewes,~J.~M.; Jovanović,~V.; Marian,~C.~M.; Dreuw,~A. On the molecular
  mechanism of non-radiative decay of nitrobenzene and the unforeseen
  challenges this simple molecule holds for electronic structure theory.
  \emph{Phys. Chem. Chem. Phys.} \textbf{2014}, \emph{16}, 12393--12406\relax
\mciteBstWouldAddEndPuncttrue
\mciteSetBstMidEndSepPunct{\mcitedefaultmidpunct}
{\mcitedefaultendpunct}{\mcitedefaultseppunct}\relax
\EndOfBibitem
\bibitem[Gray \latin{et~al.}(2017)Gray, Záliš, and Vlček]{Gray2017}
Gray,~H.~B.; Záliš,~S.; Vlček,~A. Electronic structures and photophysics of
  d8-d8 complexes. \emph{Coord. Chem. Rev.} \textbf{2017}, \emph{345},
  297--317\relax
\mciteBstWouldAddEndPuncttrue
\mciteSetBstMidEndSepPunct{\mcitedefaultmidpunct}
{\mcitedefaultendpunct}{\mcitedefaultseppunct}\relax
\EndOfBibitem
\bibitem[Haldrup \latin{et~al.}(2019)Haldrup, Levi, Biasin, Vester, Laursen,
  Beyer, Kj{\ae}r, {Brandt Van Driel}, Harlang, Dohn, Hartsock, Nelson,
  Glownia, Lemke, Christensen, Gaffney, Henriksen, M{\o}ller, and
  Nielsen]{Haldrup2019}
Haldrup,~K.; Levi,~G.; Biasin,~E.; Vester,~P.; Laursen,~M.~G.; Beyer,~F.;
  Kj{\ae}r,~K.~S.; {Brandt Van Driel},~T.; Harlang,~T.; Dohn,~A.~O.;
  Hartsock,~R.~J.; Nelson,~S.; Glownia,~J.~M.; Lemke,~H.~T.; Christensen,~M.;
  Gaffney,~K.~J.; Henriksen,~N.~E.; M{\o}ller,~K.~B.; Nielsen,~M.~M. {Ultrafast
  X-Ray Scattering Measurements of Coherent Structural Dynamics on the
  Ground-State Potential Energy Surface of a Diplatinum Molecule}. \emph{Phys.
  Rev. Lett.} \textbf{2019}, \emph{122}, 63001\relax
\mciteBstWouldAddEndPuncttrue
\mciteSetBstMidEndSepPunct{\mcitedefaultmidpunct}
{\mcitedefaultendpunct}{\mcitedefaultseppunct}\relax
\EndOfBibitem
\bibitem[Monni \latin{et~al.}(2018)Monni, Capano, Aub{\"{o}}ck, Gray, Vlček,
  and Chergui]{Monni2018}
Monni,~R.; Capano,~G.; Aub{\"{o}}ck,~G.; Gray,~H.~B.; Vlček,~A.; Chergui,~M.
  {Vibrational coherence transfer in the ultrafast intersystem crossing of a
  diplatinum complex in solution}. \emph{Proc. Natl. Acad. Sci.} \textbf{2018},
  \emph{115}, E6396--E6403\relax
\mciteBstWouldAddEndPuncttrue
\mciteSetBstMidEndSepPunct{\mcitedefaultmidpunct}
{\mcitedefaultendpunct}{\mcitedefaultseppunct}\relax
\EndOfBibitem
\bibitem[{Van Der Veen} \latin{et~al.}(2011){Van Der Veen}, Cannizzo, Mourik,
  Vl{\v{c}}ek, and Chergui]{Veen2011}
{Van Der Veen},~R.~M.; Cannizzo,~A.; Mourik,~F.~V.; Vl{\v{c}}ek,~A.;
  Chergui,~M. {Vibrational Relaxation and Intersystem Crossing of Binuclear
  Metal Complexes in Solution}. \emph{J. Am. Chem. Soc.} \textbf{2011},
  \emph{133}, 305--315\relax
\mciteBstWouldAddEndPuncttrue
\mciteSetBstMidEndSepPunct{\mcitedefaultmidpunct}
{\mcitedefaultendpunct}{\mcitedefaultseppunct}\relax
\EndOfBibitem
\bibitem[Clodfelter \latin{et~al.}(1994)Clodfelter, Doede, Brennan, Nagle,
  Bender, Turner, and Lapunzina]{Clodfelter1994}
Clodfelter,~S.~A.; Doede,~T.~M.; Brennan,~B.~A.; Nagle,~J.~K.; Bender,~D.~P.;
  Turner,~W.~A.; Lapunzina,~P.~M. Luminescent Metal-Metal-Bonded Exciplexes
  Involving Tetrakis@-diphosphito)diplatinate(II) and Thallium(1). \emph{J. Am.
  Chem. Soc.} \textbf{1994}, \emph{116}, 11379--11386\relax
\mciteBstWouldAddEndPuncttrue
\mciteSetBstMidEndSepPunct{\mcitedefaultmidpunct}
{\mcitedefaultendpunct}{\mcitedefaultseppunct}\relax
\EndOfBibitem
\bibitem[Christensen \latin{et~al.}(2010)Christensen, Haldrup, Kjær,
  Cammarata, Wulff, Bechgaard, Weihe, Harrit, and Nielsen]{Christensen2010}
Christensen,~M.; Haldrup,~K.; Kjær,~K.~S.; Cammarata,~M.; Wulff,~M.;
  Bechgaard,~K.; Weihe,~H.; Harrit,~N.~H.; Nielsen,~M.~M. Structure of a
  short-lived excited state trinuclear Ag-Pt-Pt complex in aqueous solution by
  time resolved X-ray scattering. \emph{Phys. Chem. Chem. Phys.} \textbf{2010},
  \emph{12}, 6921--6923\relax
\mciteBstWouldAddEndPuncttrue
\mciteSetBstMidEndSepPunct{\mcitedefaultmidpunct}
{\mcitedefaultendpunct}{\mcitedefaultseppunct}\relax
\EndOfBibitem
\end{mcitethebibliography}

\end{document}


\clearpage

\section{Second Hessian eigenvalue along doubly excited state energy curve of H\textsubscript{2}}
\begin{figure}
    \centering
    \includegraphics{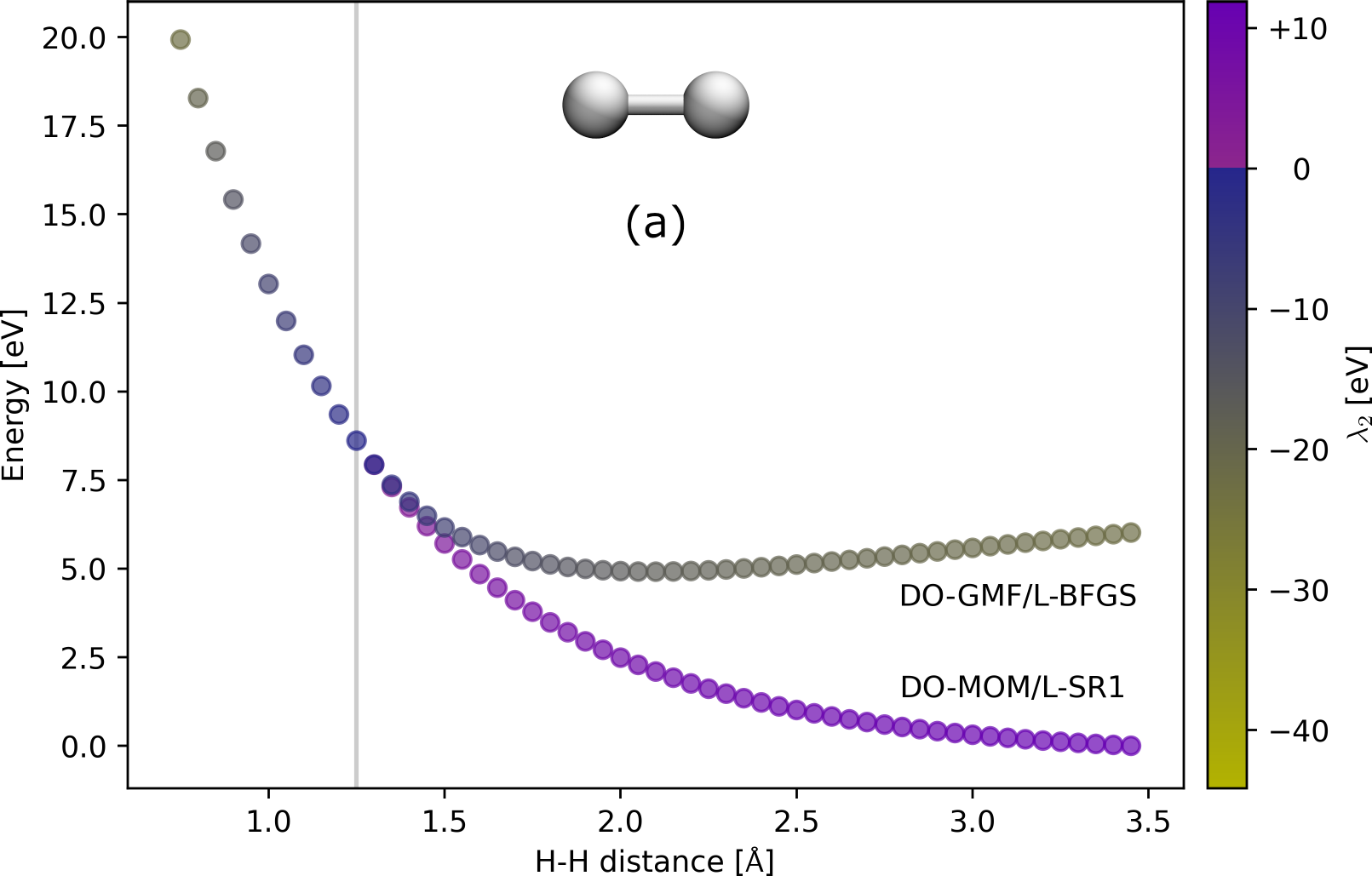}
    \caption{Calculated energy of a doubly excited state of H\textsubscript{2} as a function of the H-H distance calculated with DO-GMF and DO-MOM using sequential point acquisition. A double excitation from the ground state is generated at the first geometry at an H-H distance of 0.74\,\AA. The DO-GMF calculation targets a 2\textsuperscript{nd}-order saddle point. The points on the curves are colored according to the value of the second eigenvalue of the electronic Hessian, $\lambda_{2}$, while the gray vertical line marks where symmetry broken solutions appear. Before that, both DO-MOM and DO-GMF converge on the 2\textsuperscript{nd}-order saddle point corresponding to the symmetry-pure solution, $\sigma_g^0\sigma_u^{*2}$\,. After that, DO-MOM converges to a 1\textsuperscript{st}-order saddle point corresponding to the symmetry-pure solution giving an incorrect potential energy curve. Instead, the DO-GMF calculations keep converging on a 2\textsuperscript{nd}-order saddle point corresponding to a symmetry-broken solution with ionic character (H$^+$H$^-$/H$^-$H$^+$), thereby providing more accurate potential energy curve.}
    \label{fig:my_label}
\end{figure}
\clearpage

\section{N-Phenylpyrrole convergence}
\begin{figure}
    \centering
    \includegraphics[width=\textwidth]{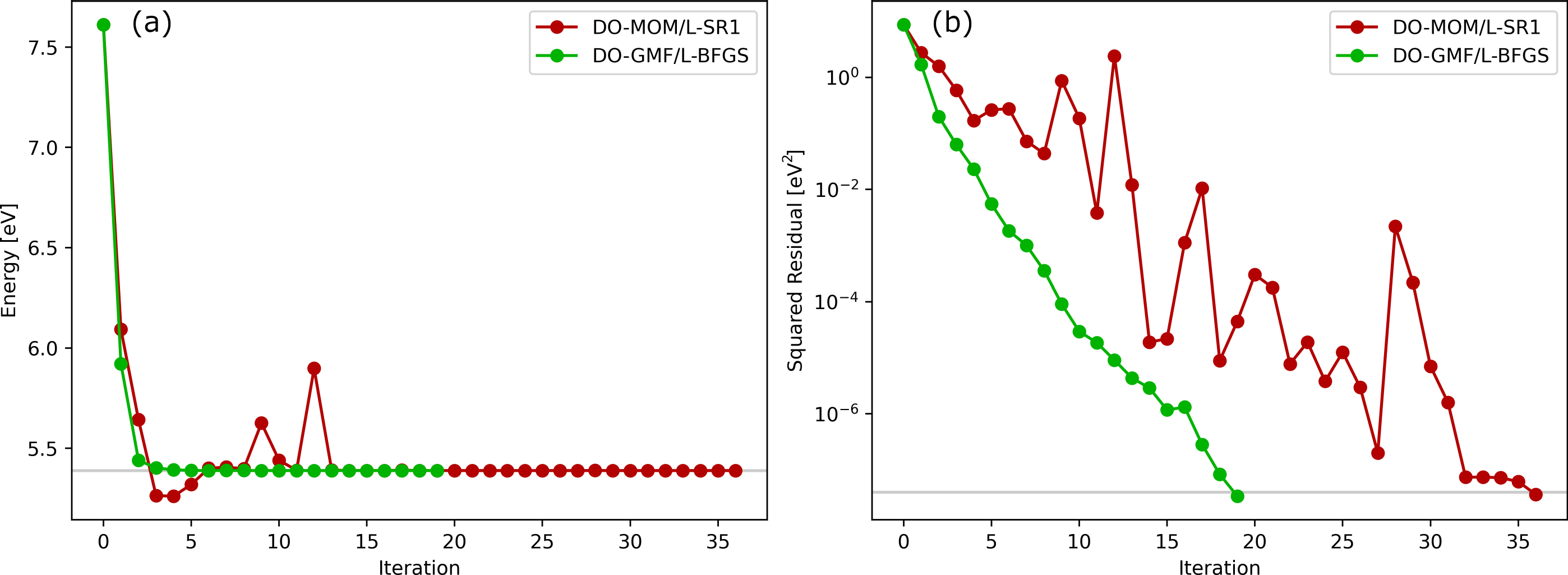}
    \caption{Convergence of the excitation energy (a) and squared residual of the KS equations (b) during DO-MOM and DO-GMF calculations of the open-shell singlet $\pi_{\mathrm{ph}}^{*} \leftarrow \pi_{\mathrm{py}}$ charge transfer excited state of nitrobenzene. Both DO-MOM and DO-GMF calculations use a maximum step length of 0.1 (in contrast to fig. 8 in the main text) for the quasi-Newton Hessian update. While DO-MOM displays erratic convergence behavior, it does converge to the target 6\textsuperscript{th}-order saddle point when this smaller optimization step size is used. DO-GMF rapidly converges to the target excited state solution.}
    \label{fig:si_PP_convergence_0.1}
\end{figure}
\clearpage

\section{Atomic configurations of all systems}
\subsection{Ethylene (optimized geometry)}
\begin{table}[!h]
    \centering
    \begin{tabular}{ c c c c }
        C & 0.66875 & 0.00000 & 0.00000 \\ 
        C & -0.66875 & 0.00000 & 0.00000 \\ 
        H & 1.24284 & 0.00000 & 0.93262 \\ 
        H & 1.24284 & 0.00000 & -0.93262 \\ 
        H & -1.24284 & 0.00000 & 0.93262 \\ 
        H & -1.24284 & 0.00000 & -0.93262
    \end{tabular}
\end{table}

\subsection{Nitrobenzene}
\begin{table}[!h]
    \centering
    \begin{tabular}{ c c c c }
        C & 8.97914 & 6.83796 & 10.66825 \\ 
        C & 10.19426 & 6.83796 & 9.99955 \\ 
        C & 10.18614 & 6.83796 & 8.61089 \\ 
        C & 8.97950 & 6.83796 & 7.91764 \\ 
        C & 7.77273 & 6.83796 & 8.61075 \\ 
        C & 7.76406 & 6.83796 & 9.99948 \\ 
        H & 11.11179 & 6.83796 & 10.56369 \\ 
        H & 11.12117 & 6.83796 & 8.07178 \\ 
        H & 8.97965 & 6.83796 & 6.83796 \\ 
        H & 6.83796 & 6.83796 & 8.07113 \\ 
        H & 6.84622 & 6.83796 & 10.56314 \\ 
        N & 8.97963 & 6.83796 & 12.14184 \\ 
        O & 7.89408 & 6.83796 & 12.71051 \\ 
        O & 10.06587 & 6.83796 & 12.70920
    \end{tabular}
\end{table}
\clearpage

\subsection{Twisted N-phenylpyrrole}
\begin{table}[h!]
    \centering
    \begin{tabular}{ c c c c }
        C & 10.13955 & 10.09711 & 12.38421 \\ 
        C & 11.34646 & 10.09711 & 13.07460 \\ 
        C & 8.93264 & 10.09711 & 13.07460 \\ 
        C & 10.13955 & 10.09711 & 15.16101 \\ 
        C & 11.34465 & 10.09711 & 14.46510 \\ 
        C & 8.93444 & 10.09711 & 14.46510 \\ 
        C & 10.13955 & 11.21630 & 10.16939 \\ 
        C & 10.13955 & 8.97792 & 10.16939 \\ 
        C & 10.13955 & 10.80762 & 8.85676 \\ 
        C & 10.13955 & 9.38661 & 8.85676 \\ 
        N & 10.13955 & 10.09711 & 10.96513 \\ 
        H & 12.26980 & 10.09711 & 12.51889 \\ 
        H & 8.00930 & 10.09711 & 12.51889 \\ 
        H & 12.27910 & 10.09711 & 15.00190 \\ 
        H & 8.00000 & 10.09711 & 15.00190 \\ 
        H & 10.13955 & 10.09711 & 16.23848 \\ 
        H & 10.13955 & 12.19423 & 10.60963 \\ 
        H & 10.13955 & 8.00000 & 10.60963 \\ 
        H & 10.13955 & 11.45309 & 8.00000 \\ 
        H & 10.13955 & 8.74113 & 8.00000
    \end{tabular}
\end{table}
\clearpage

\subsection{AgPtPOP (optimized geometry)}
\begin{table}[!h]
    \centering
    \begin{tabular}{ c c c c }
        Pt & 10.36636 & 10.36081 & 9.73625 \\ 
        Pt & 10.35958 & 10.36940 & 12.57030 \\ 
        P & 11.98825 & 12.16740 & 9.60202 \\ 
        P & 8.74412 & 8.55064 & 9.62418 \\ 
        P & 8.56487 & 11.98634 & 9.60373 \\ 
        P & 12.16350 & 8.72984 & 9.62348 \\ 
        P & 8.54921 & 11.97660 & 12.63753 \\ 
        P & 12.16615 & 8.75762 & 12.64771 \\ 
        P & 8.76220 & 8.55861 & 12.67004 \\ 
        P & 11.95954 & 12.18094 & 12.65503 \\ 
        O & 12.66786 & 12.46146 & 11.12684 \\ 
        O & 8.03785 & 8.29598 & 11.14620 \\ 
        O & 8.35874 & 12.72995 & 11.11496 \\ 
        O & 12.34246 & 7.97951 & 11.13613 \\ 
        O & 11.48354 & 13.51620 & 9.00880 \\ 
        O & 9.25295 & 7.18693 & 9.07287 \\ 
        O & 7.18877 & 11.47428 & 9.08658 \\ 
        O & 13.54876 & 9.23559 & 9.12713 \\ 
        O & 8.93818 & 13.28092 & 13.54214 \\ 
        O & 11.78224 & 7.46977 & 13.57769 \\ 
        O & 7.44285 & 8.96327 & 13.54699 \\ 
        O & 13.28943 & 11.77352 & 13.51126 \\ 
        O & 13.31964 & 11.74289 & 8.75356 \\ 
        O & 7.42811 & 8.96228 & 8.74649 \\ 
        O & 8.97261 & 13.27897 & 8.69122 \\ 
        O & 11.75869 & 7.44211 & 8.70100 \\ 
        O & 7.15935 & 11.46813 & 13.11709 \\ 
        O & 13.56270 & 9.27058 & 13.10083 \\ 
        O & 9.27375 & 7.19311 & 13.21240 \\ 
        O & 11.45425 & 13.54292 & 13.21229 \\ 
        H & 13.48616 & 10.70285 & 8.90847 \\ 
        H & 13.47770 & 10.74071 & 13.33821 \\ 
        H & 10.71750 & 7.26402 & 8.84458 \\ 
        H & 10.01789 & 13.44936 & 8.82022 \\ 
        H & 9.97789 & 13.46520 & 13.40583 \\ 
        H & 7.24855 & 9.99317 & 13.37006 \\ 
        H & 10.74369 & 7.27675 & 13.43326 \\ 
        H & 7.26078 & 10.00544 & 8.88401 \\ 
        Ag & 10.36314 & 10.36065 & 7.10520
    \end{tabular}
\end{table}